\documentclass{JFM-FLM_Au} 

\usepackage{derivative}
\usepackage{amsmath}
\usepackage{xcolor}

\newcommand{\dx}{\,\mathit{dx}}
\newcommand{\ds}{\,\mathit{ds}}
\newcommand{\dy}{\,\mathit{d}\boldsymbol{y}}
\newcommand{\nabx}{\bnabla_{x}}
\newcommand{\naby}{\bnabla_{y}}
\newcommand{\x}{\boldsymbol{x}}
\newcommand{\y}{\boldsymbol{y}}
\newcommand{\ny}{\boldsymbol{n}_y}
\newcommand{\uvec}{\boldsymbol{u}}
\newcommand{\unul}{\boldsymbol{u}_0}
\newcommand{\ujed}{\boldsymbol{u}_1}
\newcommand{\udva}{\boldsymbol{u}_2}
\newcommand{\uk}{\boldsymbol{u}_k}
\newcommand{\Jk}{\boldsymbol{J}_k}
\newcommand{\Gammav}{\mathsfbi{\Gamma}}
\newcommand{\K}{\mathsfbi{K}}
\newcommand{\Piv}{\mathsfbi{\Pi}}
\newcommand{\JGamma}{\mathsfbi{J}_\Gamma}
\newcommand{\Id}{\mathsfbi{I}_d}
\newcommand{\Cm}{C^{-}}
\newcommand{\Cp}{C^{+}}

\newcommand{\Um}{U^{-}}
\newcommand{\Up}{U^{+}}

\newcommand{\betam}{\beta^{-}}
\newcommand{\betap}{\beta^{+}}

\newcommand{\alpham}{\alpha^{-}}
\newcommand{\alphap}{\alpha^{+}}

\newcommand{\gammam}{\gamma^{-}}
\newcommand{\gammap}{\gamma^{+}}

\newcommand{\order}{\textit{O}}
\newcommand{\dv}[2]{\frac{\mathrm{d}#1}{\mathrm{d}#2}}
\newcommand{\rf}{\mathrm{ref}}

\lefttitle{V. Klika and V. Ku{\v z}el}
\righttitle{Journal of Fluid Mechanics}

\title{Coupled Transport and Adsorption in Graded Filters: A Multi-Scale Analysis of Non-Solenoidal Effects}

\author{V{\'a}clav Klika\aff{1}, Vojt{\v e}ch Ku{\v z}el\aff{1}}

\affiliation{\aff{1}Dept. Mathematics, FNSPE, Czech Technical University in Prague, Trojanova 13, 120 00 Prague, Czech Republic}

\corresau{V{\'a}clav Klika, \email{vaclav.klika@cvut.cz}}

\begin{document}
\maketitle

\begin{abstract}
We investigate the transport and adsorption of solutes within graded porous filters characterised by a spatially varying microstructure. While classical homogenisation theory typically assumes periodic media, we employ the method of multiple scales to derive an effective macroscopic model for ``near-periodic'' geometries where the porosity varies slowly over the longitudinal coordinate. A key novelty of this work is the departure from the standard solenoidal constraint; instead, we introduce a modified incompressibility condition derived from non-equilibrium thermodynamics that accounts for the coupling between the solute concentration and the solvent velocity. This leads to a generalised Darcy-scale description where the fluid velocity field is non-solenoidal within the porous domain. Through asymptotic analysis, we determine the leading-order concentration profiles and quantify first-order corrections that capture the interplay between the porosity gradient and the mixture composition. We evaluate filter performance across several metrics, including outflux concentration and total adsorption rate, under both fixed-flow and fixed-pressure-drop operating conditions. Our results demonstrate that the porosity gradient and the coupling parameter significantly influence the filtration efficiency, particularly as the medium approaches the clogging limit. The analysis reveals that the optimal filter design is highly sensitive to the chosen performance metric, highlighting the necessity of physically consistent boundary conditions and mixture dynamics in the design of high-efficiency graded filters.
\end{abstract}

\section{Introduction}
The filtration of solutes and suspended particles through porous media is a fundamental process across a broad spectrum of industrial and natural systems, ranging from groundwater remediation and reactive decontamination \citep{luckins2024effect} to the applications in pharmacology \citep{khanafer2006role} and the complex drying processes in capillary media \citep{luckins2025mathematical}. In high-performance applications, such as membrane science and microfluidic sorting, filters are increasingly engineered with spatially variable textures---known as graded filters---to optimise the trade-off between separation efficiency and hydraulic resistance. To design such devices effectively, a rigorous understanding of how microstructural heterogeneity governs macroscopic transport, phase change, and adsorption is essential \citep{battiato2019}. However, the multi-scale nature of these media often renders direct numerical simulation (DNS) of the full Navier--Stokes and Advection--Diffusion equations computationally prohibitive for design optimisation.

Asymptotic homogenisation (namely the method of multiple scales) provides a powerful framework for bridging these scales, allowing the derivation of effective macroscopic equations that retain the influence of the underlying micro-geometry \citep{mei2010homogenization}. Traditionally, these methods rely on the assumption of a locally periodic microstructure. While highly successful for uniform media, standard homogenisation is inherently limited when addressing graded filters where porosity gradients are explicitly introduced to enhance performance. Recent extensions have addressed near-periodic domains, for instance, through curvilinear coordinate transformations in biological tissues \citep{richardson2011, penta2014} or through the assumption of slowly varying properties. A significant advancement in this direction was made by \citet{bruna2015}, who derived homogenised equations for transport through obstacles with slowly varying radii. This framework was subsequently applied by \citet{dalwadi2015} to analyse how porosity gradients influence filtration efficiency and later extended to account for varying obstacle spacing \citep{auton2022} and the complex dynamics of pore-scale clogging \citep{ dalwadi2016}.

Despite these advances, most existing macroscopic filtration models assume that the fluid velocity field is solenoidal ($\nabla \cdot \mathbf{u} = 0$). While appropriate in certain situations (which we clarify below), this assumption may break down in systems where significant solute-solvent interactions or local volume changes occur during the adsorption process. In this work, we generalise the modelling framework of \citet{dalwadi2015} by incorporating a non-zero divergence in the fluid velocity, derived from the principles of non-equilibrium thermodynamics and mixture theory \citep{deGM2013non, klika2014guide, klika2019, klika2021modelling, klika2022}. This modification introduces a fundamental coupling between the chemical species transport and the momentum balance, leading to the (mixture) velocity being non-solenoidal.

The paper is organised as follows. In \S\ref{sec2}, we start by deriving the governing equations from nonequilibrum thermodynamics, namely from balance laws for a mixture of $n$ components. Then, we formulate the microscopic model for a filter composed of a graded array of obstacles. We describe the transport of solute via advection and diffusion, with adsorption modelled as a flux condition on the obstacle surfaces and the mixture velocity non-solenoid. The governing equations are upscaled in \S\ref{sec3} to yield an effective macroscopic system where microscale heterogeneities are encapsulated in effective coefficients determined by local cell problems. Next,  in \S\ref{sec4}, we simplify the geometry to an unidirectionally graded filter, derive the corresponding boundary conditions at the filter boundaries and gain analytical insight again using asymptotic methods. Finally, in \S\ref{sec5}, we solve the resulting equations for the simplified unidirectionally graded filter and examine the impact of the porosity gradient and the mixture coupling parameter on various efficiency metrics, comparing our findings with the classical solenoidal results to highlight the significance of realistic mixture dynamics in filter optimisation.

\section{Filtration in a mixture of incompressible solute and solvent} \label{sec2}
We consider a filtration model consisting of solid obstacles in a domain 
$\Omega\subset\mathbb{R}^d$, where $d\in\{2,3\}$ denotes the spatial dimension.
These obstacles may have a rather general geometry and the approach used here does not depend on their specific shape. However, to illustrate the general description, we assume that the filter is composed of $d$-dimensional balls arranged in a square (for $d=2$) or cubic (for $d=3$) lattice with spacing $\delta l$. The parameter $\delta\in\mathbb{R}^+$ is assumed to be small and positive and will later serve as the small parameter in asymptotic expansions as $\delta\to0$. The length $l$ represents the overall size of the filter.

We further assume that the radius of the balls varies with position throughout the filter, representing spatial inhomogeneity of the porous medium. The radius is described by the function $\delta l R(\tilde{\x})$, which defines the ball radius as a function of the spatial coordinate $\tilde{\x}\in\Omega$. The function $R(\tilde{\x})$ is assumed to be dimensionless. We denote the part of the space occupied by the balls as the solid domain $\Omega_s\subset\Omega$, representing the impermeable part of the filter.

We consider a mixture of fluid and solute flowing through the filter, where the solute represents the contaminant being removed. The region accessible to the fluid is denoted by the fluid domain $\Omega_f=\Omega\setminus\Omega_s$. To track the degree of contamination of the fluid, we introduce the (dimensional) solute concentration $\tilde{c}(\tilde{\x},\tilde{t})$, which depends on the spatial coordinate $\tilde{\x}$ and time $\tilde{t}$. For convenience, we extend the definition of the concentration to the entire domain by assuming $\tilde{c}(\tilde{\x},\tilde{t})=0$ for $\tilde{\x}\in\Omega_s$.

The final key feature of the model is the ability of the filter to capture solute particles representing the contaminant. This process is modelled as adsorption of solute particles onto the surfaces of the balls. For simplicity, we neglect clogging effects and assume that the geometry and volume of the solid domain $\Omega_s$ remain unchanged during the filtration process.

\subsection{Diffusion of incompressible mixture}
Let us consider a mixture of $n$ components. We start from general laws of physics and finally move on to the above specified mixture of the fluid and the solute.

For each component $k$ of the mixture, the following balance of mass holds \citep{deGM2013non, klika2014guide, klika2021modelling, klika2022}
\begin{equation} \label{eq1.1}
	\pdv{\rho_k}{t}=-\bnabla\boldsymbol {\cdot}\left(\rho_k\uk\right),
\end{equation}
where $\rho_k$ is the density of the k-th component and $\uk$ is its velocity. The right-hand side of \eqref{eq1.1} can include a source term $\hat{\rho}_{k}$ \citep{klika2014guide,klika2021modelling,klika2022}, but which we assume to be zero in our closed system, as we consider no reactions among the components. Thus, the density changes over time only due to material transport described by the term $\rho_k\uk$.

The level of description that we choose is for a mixture of components whose movement cannot be tracked individually but whose concentrations $\tilde{c}$ can be measured for any $k$, time and space.
Therefore, we consider the following state variables defining the mentioned level of description: the density of components $\rho_{k}$ and the centre of mass velocity of the mixture $\uvec$ as
\begin{equation*}
	\uvec=\sum_{k=1}^{n}\frac{\rho_k}{\rho}\uk.
\end{equation*}
We also define the total mixture density $\rho$ and the diffusion flow of the $k$-th component of the mixture as
\begin{align} \label{eq1.4}
  \rho&=\sum_{k=1}^{n}\rho_k, \\
  \Jk&=\rho_k\left(\uk-\uvec\right). 
\end{align}
Therefore, the diffusion flow $\Jk$ represents the momentum of the $k$-th component relative to the mixture's centre of mass. Summing $\Jk$ over all components yields
\begin{equation} \label{eq1.5}
\sum_{k=1}^{n}\Jk=\sum_{k=1}^{n}\rho_k\left(\uk-\uvec\right)=\sum_{k=1}^{n}\rho_k\uk-\sum_{k=1}^{n}\rho_k\uvec=0,
\end{equation}
meaning that only $n-1$ of the flows are independent. As a result, we can replace the $n$ inaccessible partial velocities $\uk$ with one state variable, the barycentric velocity $\uvec$, and $n-1$ independent diffusion fluxes $\Jk$. For the unknown diffusion fluxes $\Jk$ for which we shall seek closures in terms of the state variables $\rho_{k},~\uvec$.

We now rewrite the partial velocities $\uk$ in terms of the state variable $\uvec$ and the yet unknown diffusion fluxes $\Jk$. By expressing the velocities $\uk$ from \eqref{eq1.4} and substituting them into \eqref{eq1.1}, we obtain
\begin{equation} \label{eq1.6}
  \pdv{\rho_k}{t}=-\bnabla\boldsymbol {\cdot}\left(\rho_k\uk\right)=
  -\bnabla\boldsymbol {\cdot}\left(\Jk+\rho_k\uvec\right).
\end{equation}
By the virtue of the second law of thermodynamics, one can identify a suitable closure for the diffusion fluxes $\Jk$ in terms of the chosen state variables. As a simplest relation between the unknown fluxes $\Jk$ and state variables, we consider Fick's law \citep{deGM2013non, klika2021modelling, lebon2008understanding}
\begin{equation} \label{eq1.7}
	\Jk=-D_k\bnabla\rho_k,
\end{equation}
where $D_k$ is the diffusion coefficient of the k-th component of the mixture.
Therefore, we get the following form of the partial mass balances in terms of the state variables
\begin{equation} \label{eq1.8}
	\pdv{\rho_k}{t}=\bnabla\boldsymbol {\cdot}\left(D_k\bnabla\rho_k-\rho_k\uvec\right),
\end{equation}
which is a standard advection-diffusion equation.

\subsubsection{Fluid incompressibility}
Let us now inspect the fluid incompressibility assumption and utilise it in the governing equations.

We first define a volume fraction $\varphi_k$, which simply represents the fraction of the volume occupied by the $k$-th component with respect to the total volume. From the definition, the sum of all volume fractions satisfies the condition
\begin{equation} \label{eq1.9}
	\sum_{k=1}^{n}\varphi_k=1.
\end{equation}
We can now rewrite the density $\rho_k$ as
\begin{equation} \label{eq1.10}
	\rho_k=\varphi_k\rho_k^T,
\end{equation}
where $\rho_k^T$ is the so-called true density of the $k$-th component of the mixture. The true density is defined as a density of the stand-alone component, that is, if it were not included in a mixture. 
Then, the incompressibility of a given component $k$  means that  $\rho_k^T$ constant. Crucially, considering both the solvent and solute in our filtration problem to be incompressible does not result into the mixture being divergence free. 
That is because individual components in a mixture can rearrange in space, hence their volume fraction volumes can change \citep{klika2019}. Consequently, rewriting the partial mass balances in terms of volume fractions, we obtain
\begin{equation*}
  \pdv{\varphi_k}{t}=-\bnabla\boldsymbol {\cdot}\left(\varphi_k\uk\right).
  \end{equation*}
The incompressibility of all the components of the mixture entails
\begin{equation} \label{eq1.14}
	0=\bnabla\boldsymbol {\cdot}\left(\sum_{k=1}^{n}\varphi_k\uk\right),
\end{equation}
by virtue of \eqref{eq1.9} but which is different from vanishing divergence of (barycentric) velocity.

Rewriting partial velocities $\uk$ in terms of diffusion fluxes $\Jk$, we get after some algebra that the incompressibility condition \eqref{eq1.14} can be expressed as
\begin{equation} \label{eq1.17}
	\bnabla\boldsymbol {\cdot}\uvec=-\sum_{k=1}^{n}\frac{\bnabla\boldsymbol {\cdot}\Jk}{\rho_k^T}.
\end{equation}


With the description of a mixture of incompressible components ready, we move to the description of the filtration problem.

\subsection{Governing equations for filter}

Writing $\tilde{c}(\tilde{\x},\tilde{t})$ as the solute concentration, relabelling the diffusion coefficient to $D$ and denoting the mixture fluid velocity as $\tilde\uvec$, we get the following description of transport of solute particles within the fluid
\begin{subequations} \label{eq2.1}
	\begin{equation} \label{eq1.18}
\pdv{\tilde{c}}{\tilde{t}}=\tilde{\bnabla}\boldsymbol {\cdot}\left(D\tilde{\bnabla}\tilde{c}-\tilde{c}\tilde{\uvec}\right), \hspace{10pt} \tilde{\x}\in\Omega_f,
\end{equation}
where $\tilde{\bnabla}$ denotes derivative with respect to the coordinate $\tilde{\x}$.

Next, we represent the filtration process as adsorption of solute particles at the fluid-filter interface
	\begin{equation} \label{eq1.19}
		-\gamma\tilde{c}=\boldsymbol{n}\boldsymbol {\cdot}\left(D\tilde{\bnabla}\tilde{c}-\tilde{c}\tilde{\uvec}\right), \hspace{10pt} \tilde{\x}\in\partial\Omega_f,
	\end{equation}
\end{subequations}
where $\gamma\geq0$ is an adsorption coefficient and $\boldsymbol{n}$ is a normal vector to the boundary of the filter $\partial\Omega_f$. 

For simplicity, we neglect any changes to the filter geometry (clogging) due to the sedimentation of the solute on the surface $\partial\Omega_s$

\subsubsection{Fluid flow} \label{sec1.3.2}
To describe the fluid flow, we use the momentum balance of a stationary, incompressible fluid \citep{klika2019}
\begin{subequations} \label{eq2.2}
	\begin{equation} \label{eq2.13}
		-\tilde{\bnabla}\tilde{p}+\mu\tilde{\bnabla}^2\tilde{\uvec}=0, \hspace{10pt} \tilde{\x}\in\Omega_f.
	\end{equation}
	Here, $\tilde{p}(\tilde{\x},\tilde{t})$ stands for the pressure of the fluid and the constant $\mu$ is its viscosity.

We replace the fluid mass balance equation with the derived incompressibility condition \eqref{eq1.17}. Rewriting it for a two-component mixture using our new notation, we obtain
	\begin{equation*}
\tilde{\bnabla}\boldsymbol {\cdot}\tilde{\uvec}=\left(\frac{1}{\rho_v^T}-\frac{1}{\rho_u^T}\right)\tilde{\bnabla}\boldsymbol {\cdot}\boldsymbol{J}_u,
	\end{equation*}
	where we used $\boldsymbol{J}_v=-\boldsymbol{J}_u$ as a result of \eqref{eq1.5} for $n=2$. The index $v$ denotes the fluid (solvent) and the index $u$ denotes the solute. By substituting the assumed diffusion flux closure \eqref{eq1.7}, we obtain the final form of the incompressibility condition as	
	\begin{equation} \label{eq1.20}
\tilde{\bnabla}\boldsymbol {\cdot}\tilde{\uvec}=\left(\frac{1}{\rho_u^T}-\frac{1}{\rho_v^T}\right)\tilde{\bnabla}\boldsymbol {\cdot}\left(D\tilde{\bnabla}\tilde{c}\right), \hspace{10pt} \tilde{\x}\in\Omega_f.
	\end{equation}
	It is important to note that the right-hand side of $\eqref{eq1.20}$ is generally nonzero because we assume $\rho_u^T\neq\rho_v^T$ for two different substances. As pointed out above, this nonzero term results from the observation that for an incompressible fluid only the true densities $\rho_k^T$ are constant. On the other hand, if we assumed incompressibility simply as the constancy of the overall densities $\rho_k$, the right-hand side of \eqref{eq1.20} would be zero, derived directly from the diffusion equation \eqref{eq1.8}. 
        Thus, our approach can be viewed as a generalisation of a special case with $\tilde{\bnabla}\boldsymbol {\cdot}\tilde{\uvec}=0$, corresponding to $\rho_u^T=\rho_v^T$ , used for example in \citet{dalwadi2015}. We aim to explore in depth the effects of this non-solenoidal mixture velocity field.
	
	The final missing element is a boundary condition, which we define as the no-slip condition
	\begin{equation}
		\tilde{\uvec}=0, \hspace{10pt} \tilde{\x}\in\partial\Omega_f.
	\end{equation}

      \end{subequations}

\subsubsection{Microscopic governing equations}
The final form of the detailed governing equations are obtained after nondimensionalisation, see Appendix \ref{App.sec2.2} for more details, describing diffusion of the solute
\begin{subequations} \label{eq2.7}
	\begin{align}
		\pdv{c}{t}=\bnabla\boldsymbol {\cdot}\left(\frac{1}{\mathrm{Pe}}\bnabla c-\uvec c\right)&, \hspace{10pt} \x\in\Omega_f, \\
		-k\delta c=\boldsymbol{n}\boldsymbol {\cdot}\left(\frac{1}{\mathrm{Pe}}\bnabla c-\uvec c\right)&, \hspace{10pt} \x\in\partial\Omega_f.
	\end{align}
\end{subequations}
and flow of the solvent
\begin{subequations} \label{eq2.10}
	\begin{align}
		-\bnabla p+\delta^2\bnabla^2\uvec=0&, \hspace{10pt} \x\in\Omega_f, \\
		\bnabla\boldsymbol {\cdot}\uvec=\frac{A}{\mathrm{Pe}}\bnabla^2 c&, \hspace{10pt} \x\in\Omega_f, \label{eq2.10b} \\
		\uvec=0&, \hspace{10pt} \x\in\partial\Omega_f.
	\end{align}
      \end{subequations}
We introduced a dimensionless parameter, $A=\left(\frac{1}{\rho_u^T}-\frac{1}{\rho_v^T}\right)c_\infty$ with $c_\infty$ being a characteristic solute concentration outside of the filter. Note that the ratio $\frac{A}{\mathrm{Pe}}$ measures the deviation from the classical incompressibility condition for a single component fluid, which would lead to a zero right-hand side of \eqref{eq2.10b}, as described in Section \ref{sec1.3.2}.

\section{Effective macroscopic equations} \label{sec3}

Instead of solving the microscopic governing equations, Eqns. \eqref{eq2.7} and \eqref{eq2.10}, in a complicated microscopic geometry, we seek to find effective governing equations at a macroscopic level, where the details of the geometry are acting via the macroscopic coefficients.

To be able to derive a homogenised form of these microscopic equations, we follow the approach outlined in \citet{bruna2015,dalwadi2015,klika2023upscaling}, which allows the geometry to be almost periodic with mild spatial heterogeneity by the virtue of the method of multiple scales. For more rigorous treatment, we refer the reader to two-scale convergence approach \citep{allaire1,allaire2,allaire3}.

\begin{figure}
	\includegraphics[width=\linewidth]{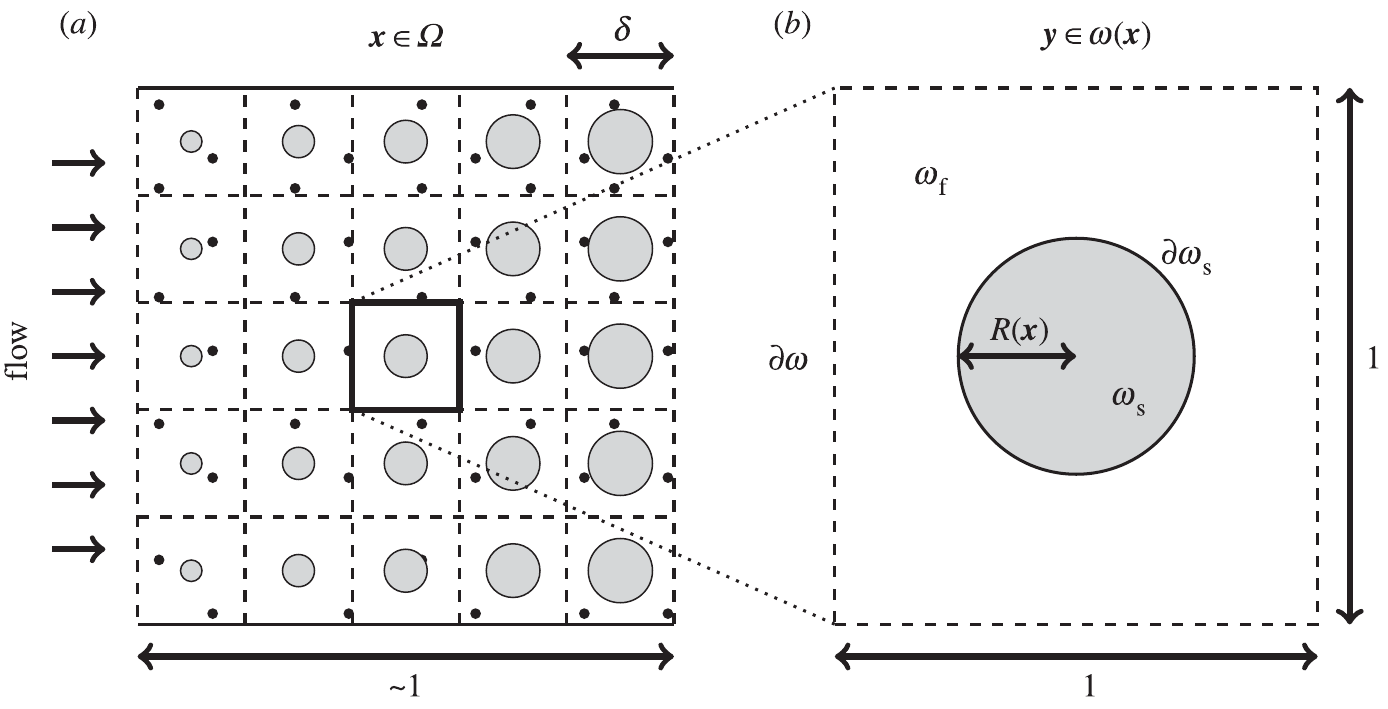}
	\caption{Diagram of the filtration model for a two-dimensional square lattice, corresponding to the model description in Section \ref{sec3.1}. (a) A macroscopic view of the entire filter, showing decreasing porosity relative to the fluid flow. (b) A microscopic close-up of an individual cell $\omega(\x)$. Source: \citet{dalwadi2015}}
	\label{fig1}
\end{figure}


\subsection{Multiple-scale analysis} \label{sec3.1}

We continue to use the variable $\x$ for the description of the filter as a whole, to which we refer as a macroscale. On the other hand, we introduce a new variable $\y=\frac{\x}{\delta}$ defined locally in a unit cell $\omega(\x)$, which will serve as a description for a microscale, see Fig. \ref{fig1}. 
Similarly to the above notation, we denote a solid part of the unit cell as $\omega_s(\x)$ and the rest of the cell accessible to the fluid as $\omega_f(\x)=\omega(\x)\setminus\omega_s(\x)$. We also denote the interface of these two regions as $\partial\omega_s(\x)$ and the boundary of the whole unit cell as $\partial\omega(\x)$. The multiple-scale analysis requires that the solution is periodic in $\y$ 
while the variables $\x$ and $\y$ are considered independent 
\citep{bender}.

Hence, we shall consider the unknowns $\uvec,c,p$ as functions of both $\x$ and $\y$. The goal now is to switch from these variables dependent on the microscale $\y$ to new effective variables dependent explicitly only on the macroscale $\x$. We define them by averaging over the unit cell $\omega(\x)$ as
\begin{subequations} \label{eq3.1real}
	\begin{align}
		C(\x,t)&=\frac{1}{|\omega(\x)|}\int_{\omega(\x)}c(\x,\y,t)\dy=\int_{\omega_f(\x)}c(\x,\y,t)\dy, \\
		\boldsymbol{U}(\x,t)&=\frac{1}{|\omega(\x)|}\int_{\omega(\x)}\uvec(\x,\y,t)\dy=\int_{\omega_f(\x)}\uvec(\x,\y,t)\dy,
	\end{align}
\end{subequations}
where we applied that $|\omega(\x)|=1$ and $c(\x,\y,t)=\uvec(\x,\y,t)=0$ for $\y\in\omega_s(\x)$. 
We also introduce a porosity $\phi$ indicating the local ratio of the available space to the total space as
\begin{equation*}
	\phi(\x)=\frac{|\omega_f(\x)|}{|\omega(\x)|}=|\omega_f(\x)|.
\end{equation*}

Further, we also introduce the fluid-volume averaged quantities, 
\begin{align} \label{eq.FluidVoluAVGquantities}
	C^f(\x,t)&=\frac{1}{|\omega_f(\x)|}\int_{\omega_f(\x)}c(\x,\y,t)\dy, \\
	\boldsymbol{U}^f(\x,t)&=\frac{1}{|\omega_f(\x)|}\int_{\omega_f(\x)}\uvec(\x,\y,t)\dy.
\end{align}
This choice seems more natural from the physical perspective as then both fields are always expressed with respect to the fluid region (and hence the presence of a filter does not change their meaning) and thus both unknowns are continuous even when porosity changes abruptly (as when entering or exiting the filter). Nevertheless, one can rewrite one in terms of the other, $C=\phi C^{f}$ and $\boldsymbol{U}=\phi \boldsymbol{U}^{f}$, and for a direct comparability with the existing literature, and in particular with \citet{dalwadi2015}, we use the former definition of the macroscopic fields in Eqns. \eqref{eq3.1real}.

\subsubsection{Transformation of derivatives} \label{sec3.1.1}
As is standard, we replace the spatial gradients as
\begin{equation} \label{eq2.14}
	\nabla\longmapsto\nabx+\frac{1}{\delta}\naby.
\end{equation}
Hence, for the fluid system \eqref{eq2.10}, we obtain after a multiplication by a suitable power of $\delta$ (so that the lowest power of $\delta$ is equal to $\delta^{0}$)
\begin{subequations} \label{eq2.16}
	\begin{align}
		-\left(\delta\nabx+\naby\right) p+\delta\left(\delta\nabx+\naby\right)^2\uvec=0&, \hspace{10pt} \y\in\omega_f(\x), \\
		\left(\delta^2\nabx+\delta\naby\right)\cdot\uvec=\frac{A}{\mathrm{Pe}}\left(\delta\nabx+\naby\right)^2 c&, \hspace{10pt} \y\in\omega_f(\x), \label{eq2.16b}\\
		\uvec=0&, \hspace{10pt} \y\in\partial\omega_s(\x). \label{eq2.16c}
	\end{align}
\end{subequations}

For the diffusion system \eqref{eq2.7} we proceed analogously, however the inhomogeneous boundary condition requires further care. We need to first describe the normal vector $\boldsymbol{n}$, perpendicular to the interface and depending on both $\x$ and $\y$. 
We now introduce a level set function describing the location of the interface, $\chi(\x,\y)=R(\x)-\|\y\|$ (hence a geometry more general than a ball is admissible), while the interface is simply $\chi(\x,\y)=0$. Now we just need to realise that the normal vector to the boundary points in the direction as follows \citep{klika2023upscaling}
\begin{equation*}
	\boldsymbol{n}\propto\nabla\chi=\left(\nabx+\frac{1}{\delta}\naby\right)\chi(\x,\y)=\nabx R(\x)-\frac{1}{\delta}\frac{\y}{\|\y\|}=\nabx R(\x)+\frac{1}{\delta}\ny,
\end{equation*}
where we defined a new unit vector on the microscale $\ny=-\frac{\y}{\|\y\|}$. The normal vector $\boldsymbol{n}$ can be then in line with above reasoning redefined as
\begin{equation*}
	\boldsymbol{n}(\x,\y)=\frac{\ny+\delta\nabx R(\x)}{\|\ny+\delta\nabx R(\x)\|}.
\end{equation*}

We may now transform the diffusion problem into the multiple-scales form which reads (after a multiplication by a suitable power of $\delta$):
\begin{subequations} \label{eq2.20}
	\begin{align}
		\delta^2\pdv{c}{t}=\left(\delta\nabx+\naby\right)\cdot\left(\frac{1}{\mathrm{Pe}}\left(\delta\nabx+\naby\right) c-\delta\uvec c\right)&, \hspace{10pt} \y\in\omega_f(\x), \\
		-k\delta^2 c=\left(\ny+\delta\nabx R(\x)\right)\cdot\left(\frac{1}{\mathrm{Pe}}\left(\delta\nabx+\naby\right) c-\delta\uvec c\right)+\order (\delta^3)&, \hspace{10pt} \y\in\partial\omega_s(\x), \label{eq2.20b}
	\end{align}
\end{subequations}
where we also multiplied equation \eqref{eq2.20b} by the term $\|\ny+\delta\nabx R(\x)\|$.


\subsubsection{Multiple-scale expansion}
We now expand the state variables  in $\delta$ into regular asymptotic expansions
\begin{align*}
	c(\x,\y,t)&=c_0(\x,\y,t)+\delta c_1(\x,\y,t)+\delta^2 c_2(\x,\y,t)+\cdots, \\
	\uvec(\x,\y,t)&=\unul(\x,\y,t)+\delta \ujed(\x,\y,t)+\delta^2 \udva(\x,\y,t)+\cdots, \\
	p(\x,\y,t)&=p_0(\x,\y,t)+\delta p_1(\x,\y,t)+\delta^2 p_2(\x,\y,t)+\cdots,
\end{align*}
which finally gives the following form of the governing equations \eqref{eq2.16} and \eqref{eq2.20}
\begin{subequations} \label{eq2.24}
	\begin{align}
		-\left(\delta\nabx+\naby\right)\left(p_0+\delta p_1+\delta^2 p_2\right)+\delta\left(\delta\nabx+\naby\right)^2\left(\unul+\delta \ujed\right)=0&, \hspace{10pt} \y\in\omega_f(\x), \\ 
		\left(\delta^2\nabx+\delta\naby\right)\cdot\left(\unul+\delta \ujed\right)=\frac{A}{\mathrm{Pe}}\left(\delta\nabx+\naby\right)^2 \left(c_0+\delta c_1+\delta^2 c_2\right)&, \hspace{10pt} \y\in\omega_f(\x), \\
		\unul+\delta \ujed+\delta^2 \udva=0&, \hspace{10pt} \y\in\partial\omega_s(\x) \label{eq2.24c}
	\end{align}
\end{subequations}
and
\begin{subequations} \label{eq2.25}
	\begin{multline}	\hspace{9pt}\delta^2\pdv{c_0}{t}=\left(\delta\nabx+\naby\right)\cdot\left(\frac{1}{\mathrm{Pe}}\left(\delta\nabx+\naby\right)\left(c_0+\delta c_1+\delta^2 c_2\right)-\delta\left(\unul+\delta \ujed\right)\left(c_0+\delta c_1\right)\right), \\
		\y\in\omega_f(\x),
	\end{multline}
	\vspace{-30pt}
	\begin{multline}
		-k\delta^2 c_0=\left(\ny+\delta\nabx R(\x)\right)\cdot\left(\frac{1}{\mathrm{Pe}}\left(\delta\nabx+\naby\right)\left(c_0+\delta c_1+\delta^2 c_2\right)-\delta\left(\unul+\delta \ujed\right)\left(c_0+\delta c_1\right)\right), \\
		\y\in\partial\omega_s(\x),
	\end{multline}
\end{subequations}
where we kept only terms up to the power of $\delta^2$. Now we put together all terms with the same power of $\delta$, which gets us equations for the leading order solutions $c_0,\unul,p_0$ and corrections $c_i,\uvec_i,p_i, \hspace{3pt} i\in\mathbb{N}$.

\subsubsection{Order-by-order solutions}
First, let us assume that $A/\mathrm{Pe}=\order(\delta^{0})$. Further, note that the no-slip fluid flow condition translates for all $i\in\mathbb{N}_0$ into the same condition
\begin{equation} \label{eq2.26}
	\uvec_i=0, \hspace{10pt} \y\in\partial\omega_s.
\end{equation}

Collecting the leading order problem $\order(\delta^{0})$ in the equations in \eqref{eq2.24} and \eqref{eq2.25}, we obtain
\begin{subequations} \label{eq2.28}
	\begin{align}
		\naby^2 c_0=0&, \hspace{10pt} \y\in\omega_f(\x), \\
		\ny\cdot\naby c_0=0&, \hspace{10pt} \y\in\partial\omega_s(\x).
	\end{align}
\end{subequations}
The solution of \eqref{eq2.28} is $c_0=c_0(\x,t)$, meaning that $c_0$ is independent of $\y$, while this solution is unique up to a constant by the virtue of the Sturm-Liouville theory.

In the same manner, we obtain the leading-order equations for \eqref{eq2.24} as 
\begin{subequations}
	\begin{align}
		\naby p_0=0&, \hspace{10pt} \y\in\omega_f(\x), \label{eq2.30a} \\
		\naby^2 c_0=0&, \hspace{10pt} \y\in\omega_f(\x). \label{eq2.30b}
	\end{align}
\end{subequations}
Equation \eqref{eq2.30b} is already satisfied by the solution $c_0=c_0(\x,t)$ of \eqref{eq2.28} and \eqref{eq2.30a} is fulfilled by $p_0=p_0(\x,t)$ being independent of $\y$. Both $c$ and $p$ are therefore independent of the microscale $\y$ in the leading order. 

We now proceed to the first-order subleading problem, that is $\order(\delta)$. Collecting terms with $\delta^1$ in \eqref{eq2.24}, and utilising $c_0=c_0(\x,t)$, yields
\begin{subequations} \label{eq2.32}
	\begin{align}
		-\naby p_1+\naby^2\unul=\nabx p_0&, \hspace{10pt} \y\in\omega_f(\x), \\
		\naby\cdot\unul=\frac{A}{\mathrm{Pe}}\naby^2 c_1&, \hspace{10pt} \y\in\omega_f(\x). \label{eq2.32b}
	\end{align}
      \end{subequations}
Similarly, the first subleading problem for solute transport reads      
\begin{subequations}
	\begin{align}
		0=\frac{1}{\mathrm{Pe}}\naby^2 c_1-c_0\naby\cdot\unul&, \hspace{10pt} \y\in\omega_f(\x), \label{eq2.34a} \\
		0=\ny\cdot\left(\frac{1}{\mathrm{Pe}}\nabx c_0 +\frac{1}{\mathrm{Pe}}\naby c_1\right)&, \hspace{10pt} \y\in\partial\omega_s(\x),
	\end{align}
      \end{subequations}
where we used the boundary condition \eqref{eq2.26} and the independence of $c_{0}$ on $y$.

We substitute $\naby\cdot\unul=\frac{A}{\mathrm{Pe}}\naby^2 c_1$ from \eqref{eq2.32b} into \eqref{eq2.34a} and get 
\begin{align*}
	\left(1-A c_0\right)\naby^2 c_1=0&, \hspace{10pt} \y\in\omega_f(\x), \\
	\ny\cdot\naby c_1=-\ny\cdot\nabx c_0&, \hspace{10pt} \y\in\partial\omega_s(\x).
\end{align*}
Note that the factor $\left(1-Ac_0\right)$ is nonzero. Recalling definitions $A=\left(\frac{1}{\rho_2^T}-\frac{1}{\rho_1^T}\right)c_\infty$ and ${\tilde{c}_0=c_\infty c_0}$ entails
\begin{equation*}
	1-Ac_0=1-\left(\frac{1}{\rho_2^T}-\frac{1}{\rho_1^T}\right)c_\infty c_0=1-\frac{\tilde{c}_0}{\rho_2^T}+\frac{\tilde{c}_0}{\rho_1^T}.
\end{equation*}
If $\tilde{c}_0=0$, then also $c_0=0$ and we simply obtain $1-Ac_0=1$. Now, let us assume that $\tilde{c}_0>0$. From the definition of the true density we have that $\rho_1^T>0$ and $\rho_2^T\geq\tilde{c}_0$, and thus $1-\frac{\tilde{c}_0}{\rho_2^T}\geq0$. All together then
\begin{equation} \label{eq2.29}
	1-Ac_0=1-\frac{\tilde{c}_0}{\rho_2^T}+\frac{\tilde{c}_0}{\rho_1^T}>0.
\end{equation}
Hence, we may simplify the first subleading diffusion problem into
\begin{subequations} \label{eq2.38}
	\begin{align}
		\naby^2 c_1=0&, \hspace{10pt} \y\in\omega_f(\x), \\
		\ny\cdot\naby c_1=-\ny\cdot\nabx c_0&, \hspace{10pt} \y\in\partial\omega_s(\x).
	\end{align}
      \end{subequations}
      
The equations \eqref{eq2.38} are linear in $c_1$ and uncoupled from fluid flow equations \eqref{eq2.32}, thus we can factorise its solution as
\begin{equation} \label{eq2.39}
	c_1(\x,\y,t)=-\nabx c_0(\x,t)\cdot\Gammav(\x,\y),
\end{equation}
where $\Gammav(\x,\y)$ is a vector satisfying
\begin{subequations} \label{eq2.40}
	\begin{align}
		\naby^2\Gamma_i=0&, \hspace{10pt} \y\in\omega_f(\x), \\
		\ny\cdot\naby\Gamma_i=n_{y,i}&, \hspace{10pt} \y\in\partial\omega_s(\x), \\
		\Gamma_i \hspace{5pt} \text{periodic}&, \hspace{10pt} \y\in\partial\omega(\x).
	\end{align}
\end{subequations}
Here notation $n_{y,i}$ denotes the i-th component of the vector $\ny$.

The above set of equations is called the cell problem and its solution $\Gammav$ is given directly by the geometry of our model. Also, let us note that in \eqref{eq2.40} the coordinate $\x$ appears only through the choice of the unit cell $\omega(\x)$. This means we can find the function $\Gammav$ for all cells simultaneously, where $x$ acts as a parameter. For further theoretical evaluation, we consider the function $\Gammav$ as already known.

Now, returning to the flow problem, the right-hand side of \eqref{eq2.32b} vanishes as  $\naby^2 c_1=0$. Thus, together with the boundary condition \eqref{eq2.26}, we obtain
\begin{subequations} \label{eq2.41}
	\begin{align}
		-\naby p_1+\naby^2\unul=\nabx p_0&, \hspace{10pt} \y\in\omega_f(\x), \\
		\naby\cdot\unul=0&, \hspace{10pt} \y\in\omega_f(\x), \\
		\unul=0&, \hspace{10pt} \y\in\partial\omega_s(\x).
	\end{align}
\end{subequations}
Equations \eqref{eq2.41} are again linear in $p_1$ and $\unul$, therefore we can also factorise the solution as
\begin{subequations}
	\begin{align}
		\unul(\x,\y,t)&=-\K(\x,\y,t)\cdot\nabx p_0(\x,t), \label{eq2.42} \\
		p_1(\x,\y,t)&=-\Piv(\x,\y,t)\cdot\nabx p_0(\x,t)+\bar{p}(\x,t),
	\end{align}
\end{subequations}
where $\K(\x,\y,t)$ is a matrix and $\Piv(\x,\y,t)$ a vector satisfying
\begin{subequations} \label{eq2.44}
	\begin{align}
		\Id-\naby\Piv+\naby^2\K=0&, \hspace{10pt} \y\in\omega_f(\x), \\
		\naby\cdot\K=0&, \hspace{10pt} \y\in\omega_f(\x), \\
		\K=0&, \hspace{10pt} \y\in\partial\omega_s(\x), \\
		\K,\Piv \hspace{5pt} \text{periodic}&, \hspace{10pt} \y\in\partial\omega(\x).
	\end{align}
\end{subequations}
Here $\Id$ denotes the identity matrix for the corresponding spatial dimension $d$, and the terms $\naby\Piv,\naby^2\K$ represents matrices and $\naby\cdot\K$ a vector defined for $i,j\in\{1,\cdots\hspace{-0.7pt},d\}$ as follows:
\begin{align*}
	\left(\naby\Piv\right)_{\bullet j}&=\naby\Piv_j, \\
	\left(\naby^2\K\right)_{ij}&=\naby^2\K_{ij}, \\
	\left(\naby\cdot\K\right)_i&=\naby\cdot\K_{\bullet i} \hspace{1pt} .
\end{align*}
As in the diffusion problem, by the factorisation we managed to split the dependence on $\x$ and $\y$. Functions $\K$ and $\Piv$ are again dependent on $\x$ only through the change of the boundary $\partial\omega_s(\x)$ and we can evaluate them for all cells $\omega(\x)$ at once. 
As with the function $\Gammav$, we for further progress assume that the functions $\K$ and $\Piv$ are known solutions to \eqref{eq2.44}.

\subsection{Macroscopic equations}
In the next step, we derive governing equations for the macroscopic quantities $C$ and $\boldsymbol{U}$ by inspecting a solvability condition for the second subleading problem.

\subsubsection{Diffusion}
For the diffusion system \eqref{eq2.25}, we continue with the asymptotic expansion to the second order in $\delta$:
\begin{subequations} \label{eq2.45}
	\begin{align}
		\pdv{c_0}{t}=\naby\cdot\left(\frac{1}{\mathrm{Pe}}\left(\nabx c_1+\naby c_2\right)-\unul c_1-\ujed c_0\right)+\qquad&\nonumber\\+\nabx\cdot\left(\frac{1}{\mathrm{Pe}}\left(\nabx c_0+\naby c_1\right)-\unul c_0\right)&, \hspace{10pt} \y\in\omega_f(\x), \label{eq2.45a} \\
		-kc_0=\ny\cdot\left(\frac{1}{\mathrm{Pe}}\left(\nabx c_1+\naby c_2\right)-\unul c_1-\ujed c_0\right)+\qquad&\nonumber \\+\nabx R(\x)\cdot\left(\frac{1}{\mathrm{Pe}}\left(\nabx c_0+\naby c_1\right)-\unul c_0\right)&, \hspace{10pt} \y\in\partial\omega_s(\x). \label{eq2.45b}
	\end{align}
\end{subequations}
Next, we integrate equation \eqref{eq2.45a} over the fluid volume $\omega_f(\x)$ as
\begin{align} \label{eq2.46}
	&\int_{\omega_f(\x)}\pdv{c_0}{t}\dy= \nonumber \\
	&\hspace{10pt}=\nabx\cdot\int_{\omega_f(\x)}\left(\frac{1}{\mathrm{Pe}}\left(\nabx c_0+\naby c_1\right)-\unul c_0\right)\dy-\int_{\partial\omega_s(\x)}\hspace{-12pt}kc_0\ds,
\end{align}
where we used the divergence theorem, the boundary condition \eqref{eq2.45b}, and the Reynolds transport theorem \citep{dalwadi2015,klika2023upscaling}
\begin{equation} \label{eq2.47}
\nabx\cdot\int_{\omega_f(\x)}\boldsymbol{v}\dy=\int_{\omega_f(\x)}\nabx\cdot\boldsymbol{v}\dy-\int_{\partial\omega_s(\x)}\nabx R(\x)\cdot\boldsymbol{v}\ds.
 \end{equation}
Note that Eqn. \eqref{eq2.46} give the same result as Fredholm alternative theorem \citep{allaire2}, which states that the system \eqref{eq2.45} for the unknown $c_2$ has a solution only if a condition given by the equation \eqref{eq2.46} is satisfied.

Now, given $c_0(\x,t)$ is independent of $\y$ and that $\int_{\omega_f(\x)}\dy=|\omega_f(\x)|=\phi(\x)$, Eqn. \eqref{eq2.46} then simplifies to
\begin{equation} \label{eq2.48}
	\phi(\x)\pdv{c_0}{t}=\nabx\cdot\left(\frac{1}{\mathrm{Pe}}\int_{\omega_f(\x)}\left(\nabx c_0+\naby c_1\right)\dy-c_0\int_{\omega_f(\x)}\unul\dy\right)-kc_0|\partial\omega_s(\x)|.
\end{equation}
If we define the Jacobi matrix $\left(\JGamma\right)_{ij}=\pdv{\Gamma_i}{y_j}, \hspace{3pt} i,j\in\{1,\cdots\hspace{-0.7pt},d\}$, we can, using \eqref{eq2.39}, write
\begin{equation*}
	\naby c_1=\naby\left(-\nabx c_0\cdot\Gammav\right)=\naby\left(-\sum_{i}\Gamma_i\pdv{c_0}{x_i}\right)=-\sum_{i}\left(\naby\Gamma_i\right)\pdv{c_0}{x_i}=-\JGamma^T\cdot\nabx c_0,
\end{equation*}
Therefore, we have another form of \eqref{eq2.48} given only in terms of $c_{0}$:
\begin{multline} \label{eq2.51}
	\phi(\x)\pdv{c_0}{t}=\nabx\cdot\left(\frac{1}{\mathrm{Pe}}\int_{\omega_f(\x)}\left(\Id-\JGamma^T\right)\cdot\nabx c_0\dy-c_0\int_{\omega_f(\x)}\unul\dy\right)-kc_0|\partial\omega_s(\x)| \\
	=\nabx\cdot\left(\frac{1}{\mathrm{Pe}}\left(\phi(\x)\Id-\int_{\omega_f(\x)}\JGamma^T\dy\right)\cdot\nabx c_0-c_0\int_{\omega_f(\x)}\unul\dy\right)-kc_0|\partial\omega_s(\x)|.
\end{multline}

Noting that $C(\x,t)\sim\phi(\x)c_0(\x,t)$ at the leading order, we can rewrite 
\eqref{eq2.51} into 
\begin{equation} \label{eq2.54}
	\pdv{C}{t}=\nabx\cdot\left(\tilde{D}\cdot\nabx C-\frac{C}{\phi}\left(\boldsymbol{U}+\tilde{D}\cdot\nabx\phi\right)\right)-fC,
\end{equation}
where $\tilde{D}$ is the effective diffusion coefficient and $f$ the effective adsorption coefficient defined as
\begin{gather}
	\tilde{D}(\phi)=\frac{1}{\mathrm{Pe}}\left(\Id-\frac{1}{\phi}\int_{\omega_f(\x)}\JGamma^T\dy\right), \label{eq2.45real} \\
	f(\phi)=k\frac{|\partial\omega_s(\x)|}{\phi}. \label{eq2.47real}
\end{gather}

\begin{figure}
	\centering
	\includegraphics[width=0.48\linewidth]{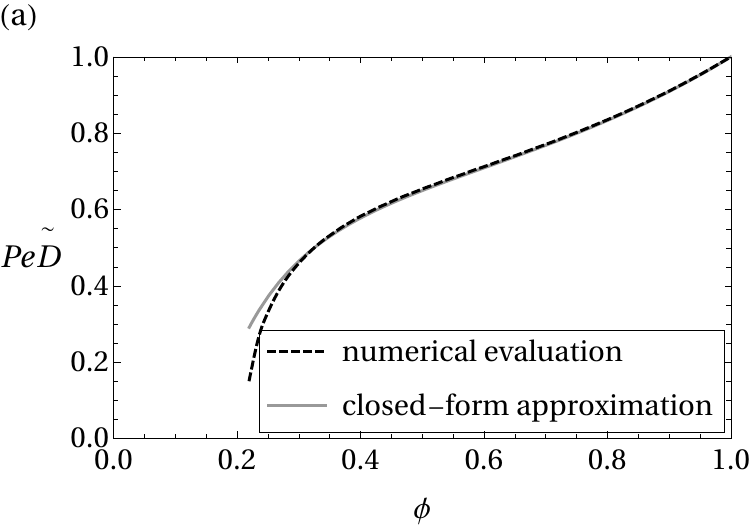}
      	\includegraphics[width=0.48\linewidth]{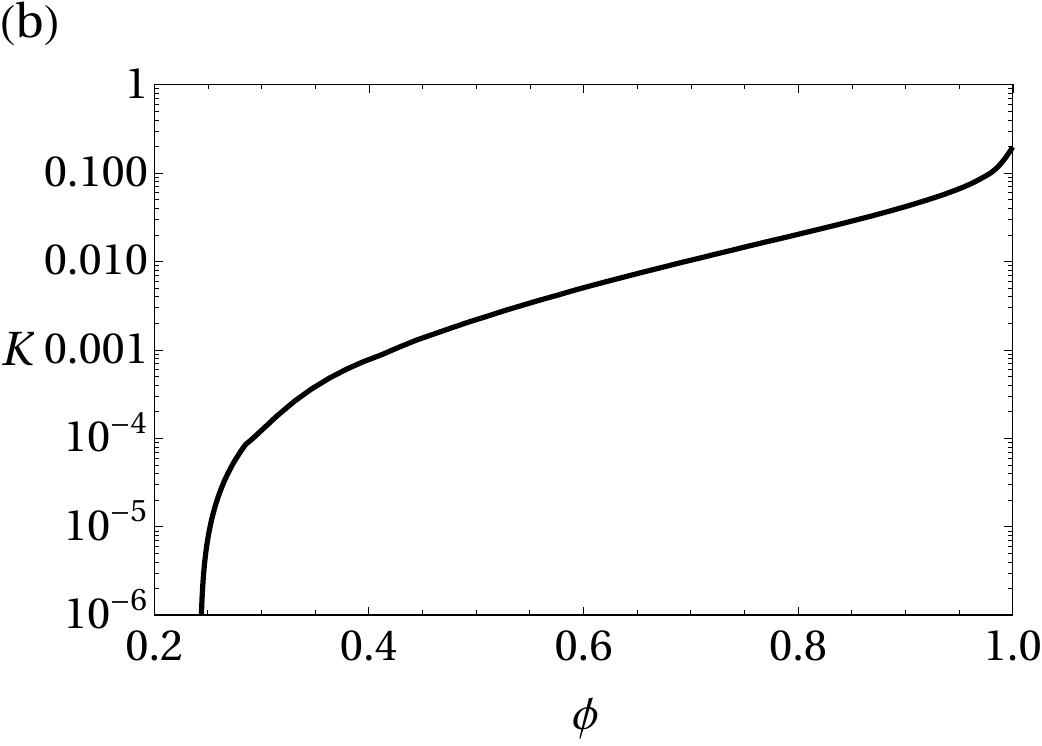}
	\caption{Evaluation of the effective macroscopic transport coefficients as a function of porosity for a two-dimensional square lattice. (left panel) Diffusion coefficient. The numerical evaluation (dashed line) is obtained by solving the cell problem \eqref{eq2.40} for various porosity values within the range $\phi\in[1-\pi/4,1]$ using the FlexPDE software. The results are then plugged into \eqref{eq2.45real} to obtain the plotted values. The closed-form approximation (solid line) represents evaluated expression \eqref{eq3.37}. (right panel) Numerical solution of the cell problem \eqref{eq2.44} yielding the effective permeability coefficient $\K=K\Id$ as a function of porosity. The numerical solution requires care and is described in Appendix \ref{sec.App-K}.}
	\label{fig2}
\end{figure}

As an illustration, we consider spherical inclusions as a representation of filter geometry, that is, by a collection of $d$-dimensional balls. In such a case, we can explicitly rewrite the function $f(\phi)$  as 
\begin{equation} \label{eq2.49}
	f(\phi)=k\frac{d(1-\phi)}{\phi}\left(\frac{V_d}{1-\phi}\right)^{\frac{1}{d}}.
\end{equation}
It is a direct consequence of the observations
\begin{equation*}
	\phi(\x)=|\omega_f(\x)|=1-|\omega_s(\x)|=1-V_dR(\x)^d \hspace{25pt}\text{and}\hspace{25pt} |\partial\omega_s(\x)|=dV_dR(\x)^{d-1},
\end{equation*}
where $V_d$ represents a volume of a unit ball in $d$ dimensions. 

We illustrate the values of $\tilde{D}(\phi)$ for spherical inclusions (filter geometry) in two dimensions in Fig. \ref{fig2}. Note that analytical approximations for several 2D packing of circles is available due to Rayleigh, see Section \ref{sec3.2} below for details.

\subsubsection{Fluid flow}
As the last step of the homogenisation process, we find macroscopic equations for the pressure $p$ and the velocity $\boldsymbol{U}$. We obtain a relation between $p$ and $\boldsymbol{U}$ at the leading order by integrating \eqref{eq2.42}:
\begin{multline*}
	\boldsymbol{U}(\x,t)\sim\int_{\omega_f(\x)}\unul(\x,\y,t)\dy=-\int_{\omega_f(\x)}\K(\x,\y,t)\cdot\nabx p_0(\x,t)\dy \\
	=-\left(\int_{\omega_f(\x)}\K(\x,\y,t)\dy\right)\cdot\nabx p_0(\x,t)\sim-K(\phi)\nabx p(\x,t),
\end{multline*}
where we defined new scalar function $K(\phi)$ as
\begin{equation} \label{eq.Ksym}
	K(\phi)\Id=\int_{\omega_f(\x)}\K(\x,\y,t)\dy,
      \end{equation}
This transcription using the identity matrix is possible due to the space symmetry of the cell problem \eqref{eq2.44}, see Appendix \ref{sec.App-K} for more details.

The macroscopic equation for the fluid flow is obtained in a similar way as the diffusion equation above.
 We have
\begin{equation} \label{eq2.61}
	\int_{\omega_f(\x)}\nabx\cdot\unul\dy=\frac{A}{\mathrm{Pe}}\int_{\omega_f(\x)}\nabx^2c_0\dy,
\end{equation}
by the virtue of $\naby c_0=0$ and $\naby\cdot\unul=0$. Next, we use the Reynolds transport theorem \eqref{eq2.47} and the boundary condition \eqref{eq2.26} to rewrite the left-hand side of \eqref{eq2.61}
\begin{equation} \label{eq2.62}
	\int_{\omega_f(\x)}\nabx\cdot\unul\dy=\nabx\cdot\int_{\omega_f(\x)}\unul\dy+\int_{\partial\omega_s(\x)}\nabx R\cdot\unul\ds\sim\nabx\cdot\boldsymbol{U}(\x,t),
\end{equation}
and $C(\x,t)\sim c_0(\x,t)\phi(\x)$ to rewrite the right-hand side
\begin{equation} \label{eq2.63}
\frac{A}{\mathrm{Pe}}\int_{\omega_f(\x)}\nabx^2c_0\dy=\frac{A}{\mathrm{Pe}}\phi(\x)\nabx^2c_0\sim\frac{A}{\mathrm{Pe}}\phi(\x)\nabx^2\left(\frac{C}{\phi(\x)}\right).
\end{equation}

Finally, by combining both \eqref{eq2.62} and \eqref{eq2.63}, we obtain the last macroscopic governing equation in the form
\begin{equation*}
	\nabx\cdot\boldsymbol{U}=\frac{A}{\mathrm{Pe}}\phi\nabx^2\left(\frac{C}{\phi}\right),
\end{equation*}
describing the fluid flow through the filter.

\subsection{Summary}
Utilising the method of multiple-scales, we obtained the following set of effective macroscopic equations for the filtration problem in terms of averaged quantities
\begin{subequations} \label{eq2.64}
	\begin{gather}
		\pdv{C}{t}=\nabx\cdot\left(\tilde{D}\cdot\nabx C-\frac{C}{\phi}\left(\boldsymbol{U}+\tilde{D}\cdot\nabx\phi\right)\right)-fC, \\
		\nabx\cdot\boldsymbol{U}=\frac{A}{\mathrm{Pe}}\phi\nabx^2\left(\frac{C}{\phi}\right), \label{eq2.64b} \\
		\boldsymbol{U}=-K\nabx p, \label{eq2.64c}
	\end{gather}
\end{subequations}
describing our filtration model at the leading order as $\delta\to 0$. The macroscopic transport parameters $\tilde{D},~K$ and the volumetric adsorption $f$ are given in terms of the microscopic domain geometry and can be evaluated separately (from the cell problem) providing an explicit link to the microscale. 

For completeness, we also list the final governing equations for the fluid-volume averaged macroscale quantity $C^{f}$ and $\boldsymbol{U}=\phi \boldsymbol{U}^{f}$ \eqref{eq.FluidVoluAVGquantities}
\begin{subequations} \label{eq2.64f}
	\begin{align}
          \pdv{C^f}{t}&=\frac{1}{\phi}\nabx\cdot\left(\phi\tilde{D}\cdot\nabx C^f-C^f\boldsymbol{U}\right)-f \, C^f, \\
          &=\frac{1}{\phi}\nabx \phi \cdot \left(\tilde{D}\cdot\nabx C^f\right)+\nabx\cdot\left(\tilde{D}\cdot\nabx C^f-C^f\boldsymbol{U}\right)-f \, C^f,
            \nonumber\\
          \nabx\cdot\boldsymbol{U}&=
                                    \frac{A}{\mathrm{Pe}}\nabx^2C^f, \\
		\boldsymbol{U}&=-K\nabx p, 
	\end{align}
\end{subequations}

Note that in the above derivation we assumed that $A/\mathrm{Pe}$ is $\order(\delta^{1})$. We leave it with the reader to check that the upscaled governing equations are in fact the same even when $A/\mathrm{Pe}$ is of lower order in $\delta$. In particular, the leading order problem is unaffected and so is the first subleading problem. 


 \section{Qualitative insight - unidirectionally graded filter} \label{sec4}

 In this section, we explore some of the consequences of the derived macroscopic governing equations \eqref{eq2.64} focusing on the nonzero term in the divergence of the mixture velocity.
 
To achieve this, we study a specific filter example with a porosity change in only one direction, which we choose to be the same as the direction of the fluid flow. This allows us to rewrite \eqref{eq2.64} only as a one-dimensional problem. We denote the only spatial dimension by variable $x$, where we assume the filter region to be $x\in(0,1)$ in accordance with the above multiple-scales asymptotics. We also assume two reservoirs, one before the filter for $x\in(-\infty,0)$ and one after the filter for $x\in(1,+\infty)$. The fluid flow is induced from $x=-\infty$ to $x=+\infty$. As the last condition, we assume the system in the steady state, meaning $\pdv{C}{t}=0$. Then, the governing equations \eqref{eq2.64} for the specified model simplify to
\begin{subequations} \label{eq3.1}
	\begin{align}
		\dv{}{x}\left(\tilde{D}\dv{C}{x}-\frac{C}{\phi}\left(U+\tilde{D}\dv{\phi}{x}\right)\right)=fC&, \hspace{10pt} x\in(0,1), \label{eq3.1a} \\
		\dv{U}{x}=\frac{A}{\mathrm{Pe}}\phi\frac{\mathrm{d}^2}{\mathrm{d} x^2}\left(\frac{C}{\phi}\right)&, \hspace{10pt} x\in(0,1). \label{eq3.1b}
	\end{align}
\end{subequations}
Because we are interested mostly in a concentration profile, we have for now omitted the equation \eqref{eq2.64c}, which connects pressure and fluid velocity. Let us also recall that the effective coefficients $\tilde{D}(\phi)$ and $f(\phi)$ are both functions of the porosity $\phi(x)$, which, for now, is an arbitrary function of~$x$, but will be specified in later analysis.

\subsection{Boundary conditions} \label{sec:BCs}
Our aim is to derive boundary conditions at the filter boundary and hence be able to solve the one-dimensional system \eqref{eq3.1} only within the filter (and not on the entire line) to obtain the desired concentration profile of the solute. To this end, we consider the filter to be located in $x\in[0,1]$  and we seek proper boundary conditions at the filter boundary, that is,  at $x\in\{0,1\}$. The system \eqref{eq3.1} consists of one second-order and one first-order differential equation, thus we need to find three boundary conditions. We first introduce boundary conditions for the whole system, including the reservoirs. Second, we extend the system \eqref{eq3.1} to both reservoirs and use its solution to determine the required boundary conditions at $x\in\{0,1\}$.

In particular, we define the upstream concentration and fluid velocity for $x\in(-\infty,0)$ as $\Cm$ and $\Um$. Analogically, we denote the downstream concentration and fluid velocity for $x\in(1,+\infty)$ as $\Cp$ and $\Up$. Two of the three boundary conditions can be defined as inlet values of concentration and fluid velocity
\begin{subequations} \label{eq3.2}
	\begin{gather}
		\Cm(x)\to 1, \hspace{10pt} x\to -\infty, \label{eq3.2a} \\
		\Um(x)\to 1, \hspace{10pt} x\to -\infty. \label{eq3.2b}
	\end{gather}
\end{subequations}
This choice of inlet values is possible due to the renormalisation in Section \ref{App.sec2.2}, where we only need to take the characteristic concentration as the average concentration of the system. The same applies also to the fluid velocity. We take the last third boundary condition as
\begin{equation} \label{eq3.3}
	\dv{\Cp}{x}\to0, \hspace{10pt} x\to +\infty,
\end{equation}
which represents an expectation that far behind the filter the concentration settles at a constant value, as in the absence of the filter there is no cause for heterogeneity. However, this final value is still unknown and must be determined as a part of the solution.

These boundary conditions correspond to experimentally controllable quantities, while the effective conditions at the filter boundary are determined by the system and cannot be imposed directly.

\subsubsection{Upstream equations}
Having the inlet and outlet conditions \eqref{eq3.2} and \eqref{eq3.3}, we can extend the system \eqref{eq3.1} to both reservoirs and find its solution. We first proceed for the upstream variables $\Cm,\Um$.
We apply limit $\phi\to 1$ to the system \eqref{eq3.1}, which yields
\begin{subequations} \label{eq3.4}
	\begin{align}
		\dv{}{x}\left(\frac{1}{\mathrm{Pe}}\dv{\Cm}{x}-\Cm\Um\right)=0&, \hspace{10pt} x\in(-\infty,0), \label{eq3.4a} \\
		\dv{\Um}{x}=\frac{A}{\mathrm{Pe}}\frac{\mathrm{d}^2\Cm}{\mathrm{d} x^2}&, \hspace{10pt} x\in(-\infty,0). \label{eq3.4b}
	\end{align}
\end{subequations}
Both equations can be integrated directly, giving
\begin{subequations} \label{eq3.9}
	\begin{align}
		\frac{1}{\mathrm{Pe}}\dv{\Cm}{x}-\Cm\Um=\betam&, \hspace{10pt} x\in(-\infty,0), \label{eq3.9a} \\
		\Um=\frac{A}{\mathrm{Pe}}\dv{\Cm}{x}+\alpham&, \hspace{10pt} x\in(-\infty,0), \label{eq3.9b}
	\end{align}
\end{subequations}
where $\alpham,\betam$ are integration constants. Substituting \eqref{eq3.9b} into \eqref{eq3.9a}, we obtain a first-order separable differential equation
\begin{equation} \label{eq3.10}
	\dv{\Cm}{x}=\mathrm{Pe}\,\frac{\betam+\alpham\Cm}{1-A\Cm}
\end{equation}
with the implicit solution
\begin{subequations} \label{eq3.7real}
	\begin{equation} \label{eq3.11}
		\alpham\left(\mathrm{Pe}~ x+\gammam\right)=\left(1+\frac{\betam}{\alpham}A\right)\ln|\betam+\alpham\Cm|-A\Cm, \hspace{10pt} x\in(-\infty,0),
	\end{equation}
	where $\gammam$ is another integration constant. Now, we take the opposite direction and substitute \eqref{eq3.10} back into \eqref{eq3.9b}, which yields the expression for the fluid velocity
	\begin{equation} \label{eq3.7b}
		\Um=\frac{\alpham+\betam A}{1-A\Cm}, \hspace{10pt} x\in(-\infty,0).
	\end{equation}
\end{subequations}

In deriving the solutions \eqref{eq3.7real}, we assumed that $\alpham\neq0$, which can be justified as follows. If we take $\alpham=0$, then by integrating \eqref{eq3.10}, we obtain a quadratic equation for $\Cm$:
\begin{equation} \label{eq3.12}
	\frac{A}{2}\left(\Cm\right)^2-\Cm+\mathrm{Pe}(\betam x + \gammam)=0.
\end{equation}
The solution to \eqref{eq3.12} is
\begin{equation*}
	\Cm(x)=\frac{1\pm\sqrt{\gamma'-2A\mathrm{Pe}\betam x}}{A}, \hspace{10pt} x\in(-\infty,0)
\end{equation*}
where we have included all constant under $\gamma'$. Now, considering the limit $x\to -\infty$, the term $(2A\mathrm{Pe}\betam x)$ diverges, which contradicts the assumption \eqref{eq3.2a}. Therefore, the only possible correction is to set ${\betam=0}$, which reduces $\Cm$ to the constant
\begin{equation*}
	\Cm=\frac{1\pm\sqrt{\gamma'}}{A}.
\end{equation*}
However, looking back at equation \eqref{eq3.9b}, the flow velocity $\Um$ would be in that case zero for the whole upstream interval $x\in(-\infty,0)$, which is also a contradiction to \eqref{eq3.2b}. Therefore, the constant $\alpham$ must be nonzero.

Moreover, we can determine the first two integration constants $\alpham,\betam$ from the inlet conditions \eqref{eq3.2}. For the limit $x\to -\infty$, the left-hand side of \eqref{eq3.11} approaches $\pm\infty$, depending on the sign of $\alpham$. However, on the right-hand side, the only term capable of going to infinity is $\ln|\betam+\alpham\Cm|$. Therefore, in order to satisfy equality for $x\to -\infty$, the following limit must apply:
\begin{equation*}
	\Cm(x)\to-\frac{\betam}{\alpham}, \hspace{10pt} x\to -\infty.
\end{equation*}
Then from the inlet concentration value \eqref{eq3.2a} we obtain
\begin{equation} \label{eq3.12real}
	\alpham=-\betam.
\end{equation}
Applying the limit $x\to -\infty$ also to the expression \eqref{eq3.7b} yields
\begin{equation*}
	1=\frac{\alpham+\betam A}{1-A},
\end{equation*}
which together with the previously derived relation \eqref{eq3.12real} gives
\begin{equation} \label{eq3.16}
	\alpham=1, \hspace{7pt} \betam=-1.
\end{equation}
These values can be substituted back to the upstream solution \eqref{eq3.7real}, giving its final form
\begin{subequations}
	\begin{align}
		\mathrm{Pe} x+\gammam=\left(1-A\right)\ln|\Cm-1|-A\Cm, \hspace{10pt} x\in(-\infty,0), \label{eq3.15a} \\
		\Um=\frac{1-A}{1-A\Cm}, \hspace{10pt} x\in(-\infty,0).
	\end{align}
\end{subequations}

Let us note that this result holds only under the assumption that $A\neq1$, which is always satisfied. In fact, $A<1$. To see this, consider the definition of $A$, the condition $A<1$ is equivalent to
\begin{equation*}
	\left(\frac{1}{\rho_2^T}-\frac{1}{\rho_1^T}\right)c_\infty<1.
\end{equation*}
Provided that $\rho_1^T,\rho_2^T,c_\infty>0$, the above expression can be rewritten as
\begin{equation} \label{eq3.22}
	1-\frac{\rho_2^T}{c_\infty}<\frac{\rho_2^T}{\rho_1^T}.
\end{equation}
We remind that the true density $\rho_2^T$ corresponds to the solute, as does the characteristic concentration $c_\infty$. The true density is defined only through the volume related to the solute, whereas the concentration is defined throughout the entire volume. Since we chose the $c_\infty$ as the average concentration within the mixture, it must hold that $c_\infty<\rho_2^T$. Therefore, the entire left-hand side of \eqref{eq3.22} is negative and the inequality is always satisfied, meaning the assumption $A<1$ is justified. Finally note, that in the limit $x\to -\infty$, the left-hand side of \eqref{eq3.15a} also approaches $-\infty$. Therefore, the term $\left(1-A\right)\ln|\Cm-1|$ must also approach minus infinity, which is satisfied as $A<1$.

\subsubsection{Downstream equations}
So far, we have specified the constants $\alpham$ and $\betam$ of the solution \eqref{eq3.7real}. To specify the third yet undetermined constant $\gammam$, we would need yet another inlet condition. However, as shown below, we do not need the value of $\gammam$ to determine the filter boundary conditions at $x\in\{0,1\}$ in order to have a well-defined problem.

In the same manner as in the upstream case, we arrive at the downstream equations
\begin{subequations}
	\begin{align}
		\dv{}{x}\left(\frac{1}{\mathrm{Pe}}\dv{\Cp}{x}-\Cp\Up\right)=0&, \hspace{10pt} x\in(1,+\infty), \label{eq3.18a} \\
		\dv{\Up}{x}=\frac{A}{\mathrm{Pe}}\frac{\mathrm{d}^2\Cp}{\mathrm{d} x^2}&, \hspace{10pt} x\in(1,+\infty).
	\end{align}
\end{subequations}
By integrating and combining them, we obtain
\begin{equation}\label{eq3.18}
	\frac{1}{\mathrm{Pe}}\dv{\Cp}{x}-\frac{A}{\mathrm{Pe}}\Cp\dv{\Cp}{x}-\alphap\Cp=\betap,
\end{equation}
where $\alphap$ and $\betap$ are integration constants defined analogously to the upstream case. Assuming that $\dv{\Cp}{x}\neq0$, solution to this equation is again in the implicit form (and same as the upstream solution)
\begin{equation} \label{eq3.19}
	\alphap\left(\mathrm{Pe} x+\gammap\right)=\left(1+\frac{\betap}{\alphap}A\right)\ln|\betap+\alphap\Cp|-A\Cp,
\end{equation}
with $\gammap$ being the last integration constant. However, we show that this solution is not admissible for the upstream problem. To see this, we consider the limit $x\to +\infty$ in equation \eqref{eq3.18}. The first two terms on the left-hand side approach zero due to $\dv{\Cp}{x}\to0$, giving the limit value for $\Cp$:
\begin{equation} \label{eq3.22real}
	\Cp(x)\to -\frac{\betap}{\alphap}, \hspace{10pt} x\to +\infty.
\end{equation}
Also applying the limit to the implicit solution \eqref{eq3.19}, we see that the left-hand side approaches $+\infty$. The positive sign is a result of the limit
\begin{equation*}
	\Up(x)\to\alphap, \hspace{10pt} x\to +\infty,
\end{equation*}
which is a consequence of $\Up=\frac{A}{\mathrm{Pe}}\dv{\Cp}{x}+\alphap$ and $\dv{\Cp}{x}\to 0$ as $x\to +\infty$. And because $\Up$ is the fluid velocity behind the filter, the value $\alphap$ must be positive.
However, on the right-hand side of \eqref{eq3.19}, the term $\ln|\betap+\alphap\Cp|$ approaches $-\infty$ as $x\to +\infty$, due to \eqref{eq3.22real}. Therefore, we must consider
\begin{equation*}
	1+\frac{\beta+}{\alpha+}A<0,
\end{equation*}
which can be rewritten as
\begin{equation*}
	\alphap<-\betap A.
\end{equation*}
Going back to the limit value of $\Cp$, we have
\begin{equation*}
	\Cp(x)\to-\frac{\betap}{\alphap}>\frac{-\betap}{-\betap A}=\frac{1}{A}>1, \hspace{10pt} x\to +\infty,
\end{equation*}
where we used the previously derived condition $A<1$. This means that the outlet concentration $\Cp$ is higher than the inlet concentration $\Cm$, which is physically incorrect, considering we have no sources of matter within the filter. Thus, the implicit solution \eqref{eq3.19} for the downstream concentration $\Cp$ is indeed not admissible.

Note, however, that the inadmissible solution was derived based on the assumption $\dv{\Cp}{x}\neq0$. Returning now to the omitted special case, we can see that the equations \eqref{eq3.9} can be solved supposing that $\Cp$ and $\Up$ are constants. To match the previous notation, we denote them as
\begin{subequations} \label{eq3.26}
	\begin{align}
		\Cp&=-\frac{\betap}{\alphap}, \\
		\Up&=\alphap.
	\end{align}
\end{subequations}
However, since these constants are outgoing values of the concentration and fluid velocity from the filter, they remain unknown and must be determined as a part of the final solution to Eqn. \eqref{eq3.1}.

\subsubsection{Final boundary conditions}
We derived the solutions \eqref{eq3.7real} and \eqref{eq3.26} for both reservoirs before and after the filter. Despite not having determined the constants $\alphap,\betap,\gamma^{\pm}$, we can use the knowledge of solutions to derive the boundary conditions for the filter itself and hence simplify the problem. For this purpose, we need a condition at the filter boundary $x\in\{0,1\}$ connecting the external variables $\Cm,\Cp,\Um,\Up$ with the internal variables $C,U$. For this, it is sufficient to take the following relations
\begin{align}
	\Cm=\frac{C}{\phi}, \; \Um=U&, \hspace{10pt} x=0, \label{eq3.27} \\
	\Cp=\frac{C}{\phi}, \; \Up=U&, \hspace{10pt} x=1, \label{eq3.28}
\end{align}
which guarantee the continuity of solute concentration and fluid velocity at the filter boundaries. From a physical point of view, we also need to impose the continuity conditions of concentration flux
\begin{align}
\frac{1}{\mathrm{Pe}}\dv{\Cm}{x}-\Cm\Um=\tilde{D}\dv{C}{x}-\frac{C}{\phi}\left(U+\tilde{D}\dv{\phi}{x}\right)&, \hspace{10pt} x=0, \label{eq3.29real} \\
\frac{1}{\mathrm{Pe}}\dv{\Cp}{x}-\Cp\Up=\tilde{D}\dv{C}{x}-\frac{C}{\phi}\left(U+\tilde{D}\dv{\phi}{x}\right)&, \hspace{10pt} x=1, \label{eq3.8d}
\end{align}
which can be derived by comparing equation \eqref{eq3.1a} with \eqref{eq3.4a} or \eqref{eq3.18a}.

Using these relations, we can finally derive the filter boundary conditions for the set of equations \eqref{eq3.1}. We take the concentration flux continuity condition \eqref{eq3.29real}, whose left-hand side is from \eqref{eq3.9a} equal to $\betam$. Knowing that $\betam=-1$, we obtain the final condition on the left filter boundary
\begin{subequations} \label{eq4.22}
	\begin{equation} \label{eq3.17}
		\tilde{D}\dv{C}{x}-\frac{C}{\phi}\left(U+\tilde{D}\dv{\phi}{x}\right)=-1, \hspace{10pt} x=0.
	\end{equation}
	Similarly for the right boundary condition, we start from equation \eqref{eq3.8d}. Considering that $\Cp$ is constant, we obtain
	\begin{equation*}
		-\Cp\Up=\tilde{D}\dv{C}{x}-\frac{C}{\phi}\left(U+\tilde{D}\dv{\phi}{x}\right), \hspace{10pt} x=1.
	\end{equation*}
	This can be further amended using \eqref{eq3.28}, yielding the final boundary condition on the right side of the filter
	\begin{equation} \label{eq3.29}
		\dv{C}{x}-\frac{C}{\phi}\dv{\phi}{x}=0, \hspace{10pt} x=1.
	\end{equation}
	
	So far, we have derived two Robin boundary conditions for the solute concentration equation \eqref{eq3.1a}. To complete the system, we also need a boundary condition for the fluid velocity equation \eqref{eq3.1b}. It can be obtained directly from the expression \eqref{eq3.7b}, using the derived constants \eqref{eq3.16} together with the relations \eqref{eq3.27} as
	\begin{equation} \label{eq3.31}
		U=\frac{1-A}{1-\frac{AC}{\phi}}, \hspace{10pt} x=0.
	\end{equation}
\end{subequations}
That the above boundary condition is well defined is clear from the definition of $C$ in \eqref{eq3.1real}, which, considering already derived observation \eqref{eq2.29}, leaves the denominator nonzero.

To summarise, we obtained the three desired boundary conditions at the filter boundary $x\in\{0,1\}$, which completes the well-posed description of the unidirectionally graded filter \eqref{eq3.1}. Moreover, this allows us to separate the filtering task from the entire region with reservoirs and limit ourselves to the solution inside the filter only.

\subsection{Slowly varying porosity} \label{sec3.3}

We now employ asymptotic methods to approximate the solution to the unidirectional filter. To this end, we consider a slowly varying porosity profile on the macroscale
\begin{equation} \label{eq3.39}
	\dv{\phi(x)}{x}=\order(\epsilon),
\end{equation}
where we introduced  a small perturbation parameter $\epsilon\ll1$. 

Next, we consider regular expansions of $C,~U$ and $\phi$ in parameter $\epsilon\to 0$ to the first order as is usual in perturbation theory:
\begin{subequations} \label{eq3.40}
	\begin{align}
		C(x)&\sim C_0(x)+\epsilon C_1(x), \\
		U(x)&\sim U_0(x)+\epsilon U_1(x), \\
		\phi(x)&\sim\phi_0+\epsilon\phi_1(x). \label{eq3.40c}
	\end{align}
\end{subequations}
Here $\phi_0$ is the mean porosity and is constant due to the assumption of the small porosity gradient \eqref{eq3.39}. 
Further, we also expand the effective coefficients $\tilde{D}$ and $f$ to the first order in $\epsilon$:
\begin{subequations} \label{eq3.41}
	\begin{align}
		\tilde{D}(x)&\sim \tilde{D}_0+\epsilon \tilde{D}_1(x), \\
		f(x)&\sim f_0+\epsilon f_1(x),
	\end{align}
      \end{subequations}
where the terms in expansions, including the observation that the leading orders $\tilde{D}_{0}$ and $f_{0}$ are independent of $x$, can be identified by the following direct calculation. We Taylor expand $\tilde{D}$ and $f$ at $\phi_{0}$:
\begin{align*}
\tilde{D}(\phi(x))&=\tilde{D}(\phi_0)+\epsilon\,\phi_1(x)\dv{\tilde{D}}{\phi}\left(\phi_0\right)+\order(\epsilon^2), \\
	f(\phi(x))&=f(\phi_0)+\epsilon\,\phi_1(x)\dv{f}{\phi}\left(\phi_0\right)+\order(\epsilon^2).
\end{align*}
Comparing with the assumed form of expansion \eqref{eq3.41} we get the desired expressions for the asymptotic terms:
\begin{align}
	\tilde{D}_0&=\tilde{D}(\phi_0), & \tilde{D}_1(x)&=\phi_1(x)\dv{\tilde{D}}{\phi}\left(\phi_0\right), \label{eq3.46real} \\
	f_0&=f(\phi_0), & f_1(x)&=\phi_1(x)\dv{f}{\phi}\left(\phi_0\right). \label{eq3.47real}
\end{align}

\subsubsection{Leading-order solution}
We substitute the expansions \eqref{eq3.40} and \eqref{eq3.41} into the set of equations \eqref{eq3.1} with boundary conditions \eqref{eq3.17}, \eqref{eq3.29} and \eqref{eq3.31} and solve order by order in $\epsilon$.
For the leading order, we obtain the equations
\begin{subequations} \label{eq3.46}
	\begin{align}
		\dv{}{x}\left(\tilde{D_0} \dv{C_0}{x}-\frac{C_0}{\phi_0}U_0\right)=f_0 C_0, \hspace{10pt} x\in(0,1), \\
		\dv{U_0}{x}=\frac{A}{\mathrm{Pe}}\frac{\mathrm{d}^2 C_0}{\mathrm{d} x^2}, \hspace{10pt} x\in(0,1) \label{eq3.46b}
	\end{align}
\end{subequations}
with the respective boundary conditions
\begin{subequations}
	\begin{align}
		\tilde{D_0}\dv{C_0}{x}-\frac{C_0}{\phi_0}U_0=-1, \hspace{10pt} x=0, \\
		U_0=\frac{1-A}{1-\frac{A C_0}{\phi_0}}, \hspace{10pt} x=0, \label{eq3.47b} \\
		\dv{C_0}{x}=0, \hspace{10pt} x=1.
	\end{align}
\end{subequations}
After some algebra, 
the set of equations \eqref{eq3.46} leads to a second-order differential equation with polynomial coefficients. A solution to such an equation is in the form of hypergeometric functions and is therefore not suitable for our purpose to obtain a closed form solution, in which higher-order solutions are expressed in terms of the lower orders.

Therefore, to progress further, we resort to one further simplification, that is, the constant $A$ is of order $\order(\epsilon)$, specifically
\begin{equation*}
	A= \epsilon \bar{A}.
      \end{equation*}
Note that as $A=\left(\frac{1}{\rho_2^T}-\frac{1}{\rho_1^T}\right)c_\infty$, its smallness is justified for example when the solution is dilute and/or the true concentrations of the solute and solvent are not different by orders of magnitude.

Equation \eqref{eq3.46b} with boundary condition \eqref{eq3.47b} then decouples into
\begin{subequations}
	\begin{align}
		\dv{U_0}{x}=0&, \hspace{10pt} x\in(0,1), \label{eq3.49a} \\
		U_0=1&, \hspace{10pt} x=0 \label{eq3.49b}
	\end{align}
\end{subequations}
with the simple solution
\begin{equation*}
	U_0=1, \hspace{10pt} x\in (0,1).
\end{equation*}
The rest of the equations then form a second separate system
\begin{subequations}
	\begin{align}
		\tilde{D_0}\frac{\mathrm{d}^2 C_0}{\mathrm{d} x^2}-\frac{1}{\phi_0}\dv{C_0}{x}-f_0 C_0=0&, \hspace{10pt} x\in(0,1), \label{eq3.51a} \\
		\tilde{D_0}\dv{C_0}{x}-\frac{C_0}{\phi_0}=-1&, \hspace{10pt} x=0, \label{eq3.51b} \\
		\dv{C_0}{x}=0&, \hspace{10pt} x=1, \label{eq3.51c}
	\end{align}
\end{subequations}
being a second-order differential equation with constant coefficients. Therefore, the leading-order solution reads
\begin{equation} \label{eq4.63}
	C_0(x)=2\alpha\phi_0 e^{ax}\left[b\cosh(ab(1-x))+\sinh(ab(1-x))\right],
\end{equation}
 where
\begin{equation*}
	a=\frac{1}{2\tilde{D_0}\phi_0}, \hspace{5pt} b=\sqrt{1+\frac{f_0}{a^2 \tilde{D_0}}},
\end{equation*}
and
\begin{equation*}
	\alpha=\frac{1}{(b^2+1)\sinh(ab)+2b\cosh(ab)}.
\end{equation*}


\subsubsection{First-order solution}
The leading-order solution corresponds to the case when $A=0$ due to our simplifying assumption $A=\order(\epsilon)$. Therefore, the first-order solution represents a correction due to the effect of mixture of two incompressible components.

Again, substituting the expansions  \eqref{eq3.40} and \eqref{eq3.41} into the set \eqref{eq3.1} with boundary conditions \eqref{eq3.17}, \eqref{eq3.29}, \eqref{eq3.31} and equating terms of order $\order(\epsilon)$, we obtain
\begin{subequations}
	\begin{align}
		\tilde{D_0}\frac{\mathrm{d}^2 C_1}{\mathrm{d} x^2}-\frac{1}{\phi_0}\dv{C_1}{x}-f_0 C_1=-\dv{}{x}\left(\tilde{D_1}\dv{C_0}{x}-\frac{C_0}{\phi_0}\left(U_1+\tilde{D_0}\dv{\phi_1}{x}-\frac{\phi_1}{\phi_0}\right)\right)+f_1 C_0&, \hspace{10pt} x\in(0,1), \label{eq3.60a} \\
		\dv{U_1}{x}=\frac{\bar{A}}{\mathrm{Pe}}\frac{\mathrm{d}^2 C_0}{\mathrm{d} x^2}&, \hspace{10pt} x\in(0,1) \label{eq3.60b}.
	\end{align}
\end{subequations}
The boundary conditions are then
\begin{subequations}
	\begin{align}
		\tilde{D_0} \dv{C_1}{x}-\frac{C_1}{\phi_0}=-\tilde{D_1}\dv{C_0}{x}+\frac{C_0}{\phi_0}\left(U_1+\tilde{D_0}\dv{\phi_1}{x}-\frac{\phi_1}{\phi_0}\right)&, \hspace{10pt} x=0, \label{eq3.61a}\\
		U_1=\bar{A}\left(\frac{C_0}{\phi_0}-1\right)&, \hspace{10pt} x=0, \label{eq3.61b} \\
		\dv{C_1}{x}=\frac{C_0}{\phi_0}\dv{\phi_1}{x}&, \hspace{10pt} x=1. \label{eq3.61c}
	\end{align}
\end{subequations}
The flow equation \eqref{eq3.60b} can be directly integrated, giving the solution
\begin{equation} \label{eq4.66}
	U_1(x)=\frac{\bar{A}}{\mathrm{Pe}}2\alpha\phi_0 a\left(1-b^2\right)e^{ax}\sinh(ab(1-x))+\theta,
\end{equation}
where $\theta$ follows from the boundary condition \eqref{eq3.61b}, yielding
\begin{equation*}
	\theta=\bar{A}\left(\left(1-\frac{\phi_0}{\mathrm{Pe}}a\left(1-b^2\right)\right)2\alpha\sinh(ab)+2\alpha b\cosh(ab)-1\right).
      \end{equation*}
      
      We note that with the velocity $U_1$ determined, all the terms on the right-hand side of \eqref{eq3.60a} are explicit known functions of space $x$. Therefore, we are left with an inhomogeneous second-order differential equation with constant coefficients. The solution can be readily found using the method of variation of parameters and reads
\begin{equation} \label{eq3.72}
	C_1(x)=\frac{e^{ax}}{2ab\tilde{D_0}}\left[\left(\int G(x)e^{-(a+ab)x}\dx+\gamma_1\right)e^{abx}-\left(\int G(x)e^{-(a-ab)x}\dx+\gamma_2\right)e^{-abx}\right],
\end{equation}
where $G(x)$ is defined as the right-hand side of \eqref{eq3.60a}. Further, the constants satisfy
\begin{multline*}
	\gamma_1=\frac{\alpha}{2}\left(-(b+1)^2 e^{ab}\left(\beta_1(1)-\beta_1(0)\right)-\frac{(b+1)^3}{b-1}e^{ab}\beta_2(0)+\left(b^2-1\right)e^{-ab}\beta_2(1) \right. \\
	\left. +2\alpha b\frac{b+1}{a}\dv{\phi_1}{x}\left(1\right)-2\frac{(b+1)^2}{b-1}\phi_0 e^{ab}N(0)\right)+\frac{b+1}{b-1}\beta_2(0)-\beta_1(0)+2\frac{\phi_0}{b-1}N(0),
\end{multline*}
\vspace{-17pt}
\begin{multline*}
	\gamma_2=\frac{\alpha}{2}\,\Bigg(\hspace{-2.5pt}-\left(b^2-1\right) e^{ab}\left(\beta_1(1)-\beta_1(0)\right)-(b+1)^2e^{ab}\beta_2(0)+(b-1)^2e^{-ab}\beta_2(1) \\
	+2\alpha b\frac{b-1}{a}\dv{\phi_1}{x}\left(1\right)-2(b+1)\phi_0 e^{ab}N(0)\Bigg),
\end{multline*}
with
\begin{align*}
	\beta_1(x)=\frac{1}{2ab\tilde{D_0}}\int G(x)e^{-(a+ab)x}\dx + \gamma_1, \\
	\beta_2(x)=-\frac{1}{2ab\tilde{D_0}}\int G(x)e^{-(a-ab)x}\dx + \gamma_2,
\end{align*}
and where we denoted the function $N(x)$ as the right-hand side of \eqref{eq3.61a}.

\subsection{Numerical solution} \label{sec3.2}
In this section, we solve the system \eqref{eq3.1} with the derived boundary conditions \eqref{eq4.22} numerically.
To this end, the governing equations were rewritten as a system of first-order ordinary differential equations by introducing auxiliary variables for the derivatives. The resulting nonlinear boundary value problem was solved in MATLAB using the built-in solver bvp4c.
Parameter sweeps were carried out using continuation, with each converged solution used as the initial guess for the next parameter value.

For the solution, we use a family of linear porosity profile parametrised as
\begin{equation} \label{eq3.32}
	\phi(x)=\phi_0+m(x-0.5), \hspace{10pt} x\in (0,1),
\end{equation}
where $\phi_0$ is the mean porosity within the filter and $m$ represents the degree of the porosity gradient. For example, for $m<0$ is the porosity gradient negative, meaning the filter obstacle size is smallest at the entrance of the filter and increases in size with increasing $x$. The value $m=0$ then represents a uniform filter with constant porosity.

Finally, note that unless we want to assess pressure, there is no need to calculate the permeability coefficient $K$ from the cell problem \eqref{eq2.44} as the unknown pressure field $p$ decouples from the two macroscopic quantities $\boldsymbol{U}, C$, see eqn \eqref{eq2.64}.

\subsubsection{Square lattice}
To evaluate the solution, we also need to specify the effective coefficients $\tilde{D}(\phi)$ and $f(\phi)$ which, as pointed out above, are dependent on the filter microscale geometry. The adsorption coefficient $f(\phi)$ is for spherical inclusions (as a filter geometry) in two dimensions from \eqref{eq2.49} equal to
\begin{equation} \label{eq3.36}
	f(\phi)=k\frac{2\pi}{\phi}\sqrt{\frac{1-\phi}{\pi}}.
\end{equation}
For the diffusion coefficient $\tilde{D}(\phi)$, we start from closed-form approximation for square lattice derived in \citet{johannesson}:
\begin{equation} \label{eq3.37}
	\tilde{D}(\phi)=\frac{1}{\mathrm{Pe}\,\phi}\left(1-\frac{2(1-\phi)}{2-\phi-0.3058(1-\phi)^4}\right), \hspace{10pt} \phi>0.3,
\end{equation}
where the volume fraction value on the right indicates the values for which the approximation is valid.
 As shown in Fig. \ref{fig2}, the approximation provides very good results for porosity $\phi>0.3$.


\begin{figure}
	\centering
	\includegraphics[width=0.48\linewidth]{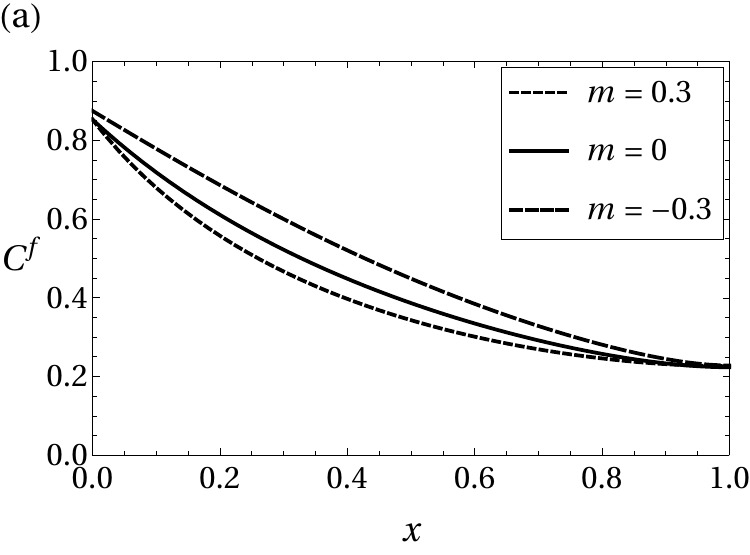} 	\includegraphics[width=0.48\linewidth]{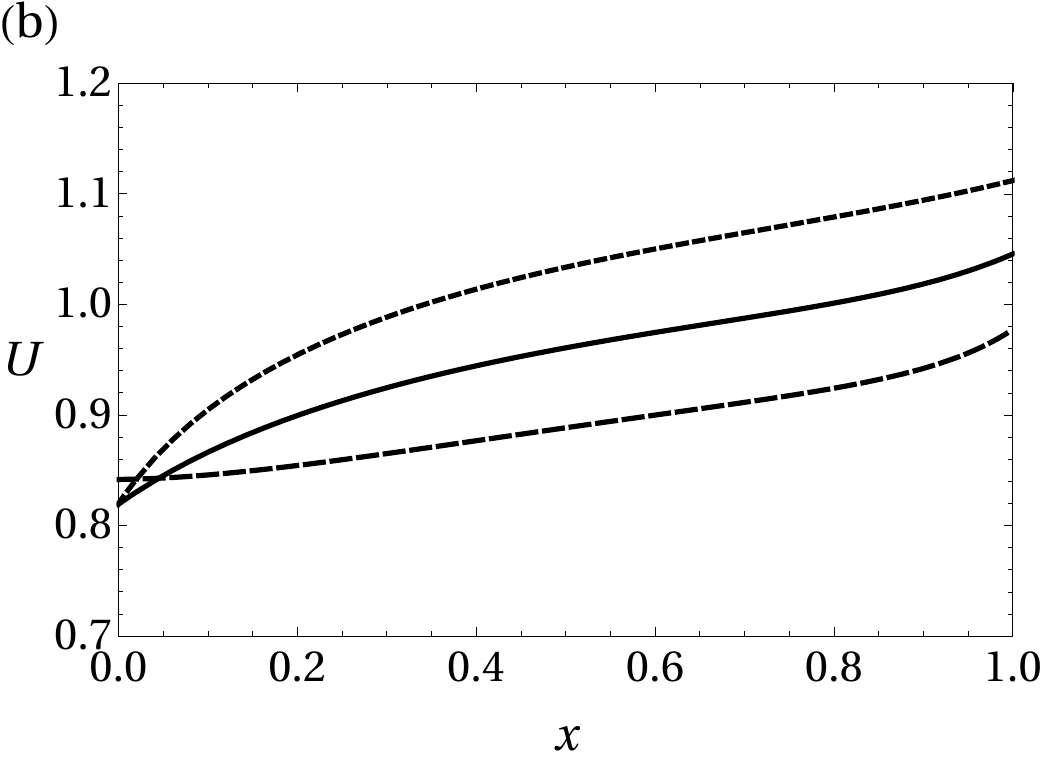}
	\caption{ Numerically obtained fluid-volume averaged solute concentration $C^{f}=C/\phi$ (a), fluid velocity $U=\phi U^{f}$ (b) 
          for three gradients $m=-0.3,0,0.3$. The evaluation is performed for a two-dimensional square lattice with the parameters $\phi_0=0.75,\, A=0.6,\, \mathrm{Pe}=3$ and $k=1$. Note that for the special case of divergence free flow field, $A=0$, the solution is $U=1$ and hence it can be viewed as a measure of the effect of the novel term, being a result of treating the filtration problem as a true mixture.}
	\label{fig3}
      \end{figure}

Using the relations for the effective macroscale coefficients, we evaluate the solutions $C^{f}$ and $U$ for various values of parameters $A,~\mathrm{Pe},~k$ and $m$ in Fig. \ref{fig3}. We opt the fluid-volume averaged quantity $C^{f}$, which needs to be monotonic and flatten out to zero derivative when leaving the filter. The fluid flow velocity, on the other hand, is chosen to be $U=\phi U^{f}$ as it allows a simple comparison to the special case $A=0$ (fluid flow being divergence free), where the solution is known to be $U=1$, see below. 
As expected, the concentration value decreases with increasing $x$, as the solute particles are being filtered out throughout the filter. We also observe that with increasing porosity gradient, the rate of concentration change decreases. This behaviour is also to be expected, because as the porosity gradient increases, more contaminant is filtered out towards the front of the filter, where the filter obstacle size is larger. Therefore, the concentration near the front should decrease faster. The same logic applies to the end of the filter, where 
the overall concentration profile becomes more uniform. Note that such observations hold for $C^{f}$, not $C$.

In addition, note that the solute concentration $C^{f}$ does not enter the filter with the (normalised) free-flow upstream value, that is $C^{f}(x=0)\neq 1$. The reason is that, as we have shown above in Section \ref{sec:BCs}, the experimentally controlled influx conditions at $x=-\infty$ translate into nontrivial Robin boundary conditions at the entry to the filter $x=0$, Eqn. \eqref{eq3.17}. As can be seen from the boundary condition \eqref{eq3.17} and also from the numerical solutions in Fig. \ref{fig3}, the filter presence and geometry significantly affects the composition of the mixture even \emph{before (upstream) the filter} and entails various entry values of the solute concentration.

\begin{figure}
	\centering
	\includegraphics[width=0.65\linewidth]{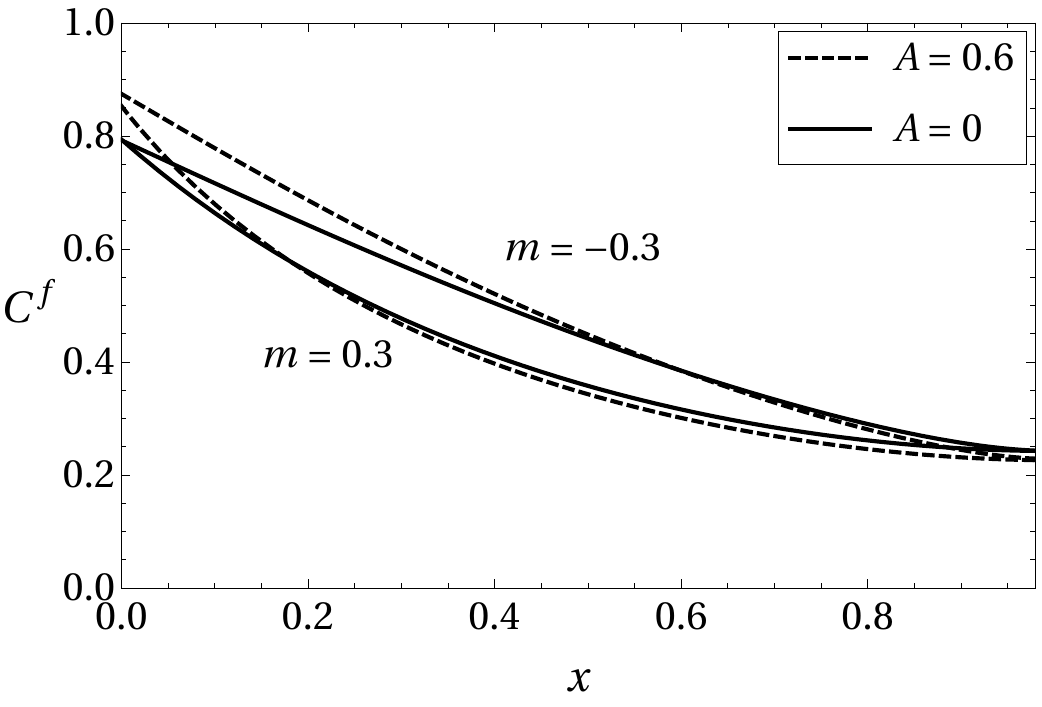}
	\caption{
          Numerically obtained concentration profiles for linear porosity \eqref{eq3.32} for $A=0$ and $A=0.6$, hence highlighting the effect of the new term. The evaluation is performed for a two-dimensional square lattice with the parameters ${\phi_0=0.75,\, \mathrm{Pe}=3,\, k=1}$ and $m=-0.3,0.3$. The lower two curves (solid and dashed) correspond to positive porosity gradient $m=0.3$, while the upper two to a negative porosity gradient $m=-0.3$. Note that the inlet solute concentration is affected (being larger for $A\neq 0$) as well as the amount of the particles being filtered out (again, being larger for $A\neq 0$). }
	\label{fig4}
\end{figure}

To examine the effect of the nonzero mixture velocity divergence, that is the right-hand side of \eqref{eq3.1b}, we already mentioned that we plot $U$ instead of $U^{f}$ Fig. \ref{fig3}b. The velocity profile is affected significantly, as for $A=0$ the solution from \eqref{eq3.1b} and \eqref{eq3.31} is $U=1$, regardless of the used porosity profile. In addition, we calculate and plot the effect on the concentration profile, see Fig. \ref{fig4}. We observe that increasing $A$ leads to a more pronounced concentration change during the filtration. 
Therefore, this represents the first major qualitatively different result of our filtration model.

\subsubsection{Hexagonal lattice}
Until now, we have only examined the filtration process in dependence on various porosity profiles $\phi(x)$ and parameters $A,\mathrm{Pe},k$. However, the effective coefficients $D(\phi),f(\phi)$ from their definitions \eqref{eq2.45real}, \eqref{eq2.47real} depend not only on the porosity $\phi$, but also on the geometry of the filter. This can be seen from the cell problem \eqref{eq2.40} and constitutes one of the advantages of asymptotic approach to deriving the macroscale equations, as the method provides a link between microscale and macroscale description. The solution to the cell problem explicitly reflects the geometry of the unit cell $\omega$. To demonstrate this effect and link to the microscale, we introduce a new filter layout in the form of a hexagonal lattice in two dimensions (with the same spherical inclusions). We represent the hexagonal lattice by unit cells $\omega=\left[-3^{-1/4},3^{-1/4}\right]\times\left[-3^{1/4},3^{1/4}\right]$ with one solid disk placed in the centre and a quarter of a disk placed in each corner of the cell. By this definition, we achieve an uniform distribution of the disks, while keeping the cell size at $|\omega|=1$.

To evaluate the solution for such layout, we first need to modify the effective coefficients $D,f$ for this specific geometry. Since every hexagonal cell contains two disks in total, we can rewrite the adsorption coefficient analogous to \eqref{eq2.49} as
\begin{equation*}
	f_{\mathrm{H}}(\phi)=k\frac{4\pi}{\phi}\sqrt{\frac{1-\phi}{2\pi}},
\end{equation*}
where the index H denotes the hexagonal lattice. Next, we take the diffusion coefficient from \citet{johannesson} using the same steps as for the square lattice in \eqref{eq3.37}, yielding
\begin{equation*}	\tilde{D}_{\mathrm{H}}(\phi)=\frac{1}{\mathrm{Pe}\,\phi}\left(1-\frac{2(1-\phi)}{2-\phi-0.07542(1-\phi)^6}\right), \hspace{10pt} \phi>0.2.
\end{equation*}

\begin{figure}
	\centering
	\includegraphics[width=0.65\linewidth]{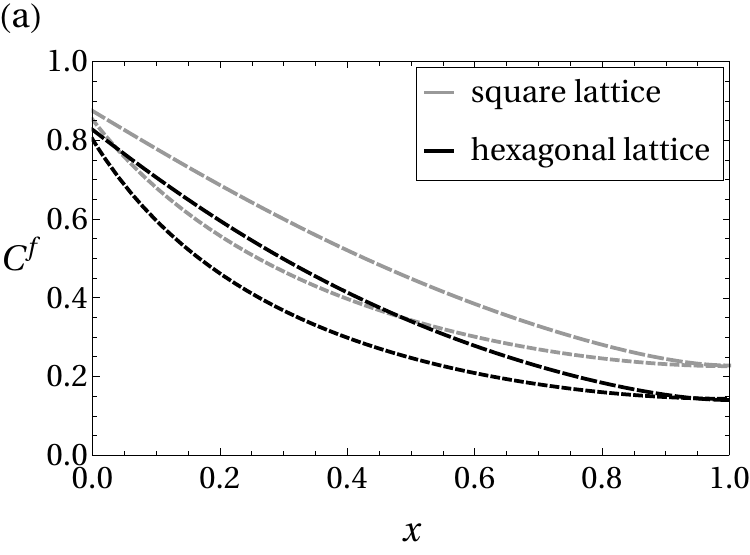}
	\caption{Numerically obtained concentration profiles for linear porosity \eqref{eq3.32} for square and hexagonal lattices in two dimensions. The parameters used are $\phi_0=0.75$, $A=0.6,\, \mathrm{Pe}=3,\, k=1$ and $m=-0.3,0.3$. The effect of microscale geometry and the capability of the asymptotic method to reflect these effects in the derived macroscale model is apparent.}
	\label{fig5}
\end{figure}

Having specified the above coefficients, we evaluate concentration profiles for the linear porosity \eqref{eq3.32} in Fig. \ref{fig5}. As expected, we observe slightly different results for both lattices, despite using the same porosity values. Specifically, we see that the hexagonal layout causes higher solute adsorption at the front of the filter. This may be a consequence of a more uniform distribution of the balls (disks) within the filter, in comparison to the square lattice. However, to confirm this phenomenon, further analysis would be required.

\subsubsection{Asymptotic approximation}
Taking the advantage of having both numerical and approximate solution, we check their validity by mutual comparison. In particular, we compare the asymptotic approximation of the solution for slowly varying porosity with the numerical results obtained above. 

For this purpose, we once again use the liner porosity profile $\phi(x)=\phi_0+m(x-0.5)$. Comparing it with the asymptotic expansion \eqref{eq3.40c}, we require
\begin{equation*}
	m=\epsilon, \hspace{10pt} \phi_1(x)=x-0.5.
\end{equation*}
We also need to specify the effective coefficients $\tilde{D}_0,\tilde{D}_1$ and $f_0,f_1$. The adsorption coefficients $f_0$ and $f_1$ can be determined from prescription \eqref{eq3.47real} using previously derived value \eqref{eq3.36} as
\begin{align*}
	f_0&=k\frac{2\pi}{\phi_0}\sqrt{\frac{1-\phi_0}{\pi}}, \\
	f_1(x)&=k\phi_1(x)\frac{\phi_0-2}{\phi_0^2}\sqrt{\frac{\pi}{1-\phi_0}}.
\end{align*}
In the same manner, we obtain the diffusion coefficients from \eqref{eq3.46real} using \eqref{eq3.37} as
\begin{align*}
	\tilde{D}_0&=\frac{1}{\phi_0}\left(1-\frac{2(1-\phi_0)}{2-\phi_0-0.3058(1-\phi_0)^4}\right), \\
	\tilde{D}_1(x)&=\frac{\phi_1(x)}{\mathrm{Pe}\,\phi_0^2}\left(-1+\frac{2(1-\phi_0)}{2-\phi_0-0.3058(1-\phi_0)^4}+\frac{\left(2+1.8348(1-\phi_0)^4\right)\phi_0}{\left(2-\phi_0-0.3058(1-\phi_0)^4\right)^2}\right).
\end{align*}

\begin{figure}
	\includegraphics[width=0.49\linewidth]{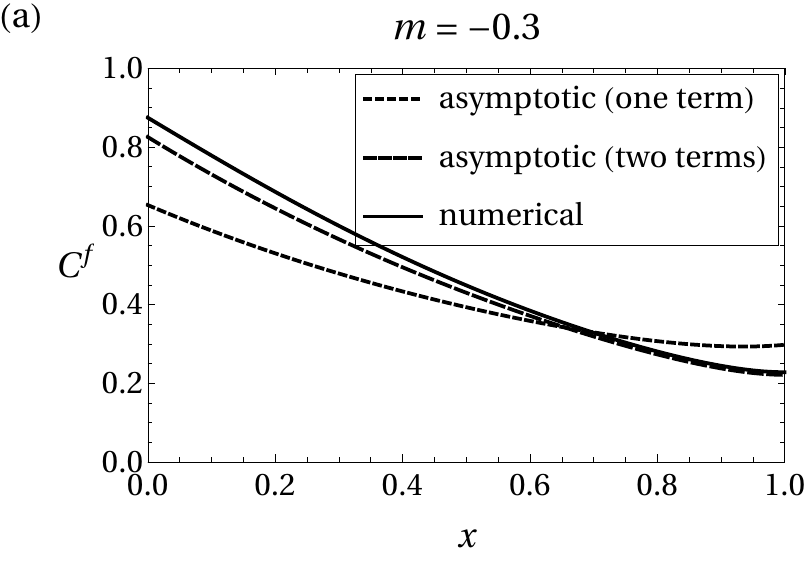}
	\includegraphics[width=0.49\linewidth]{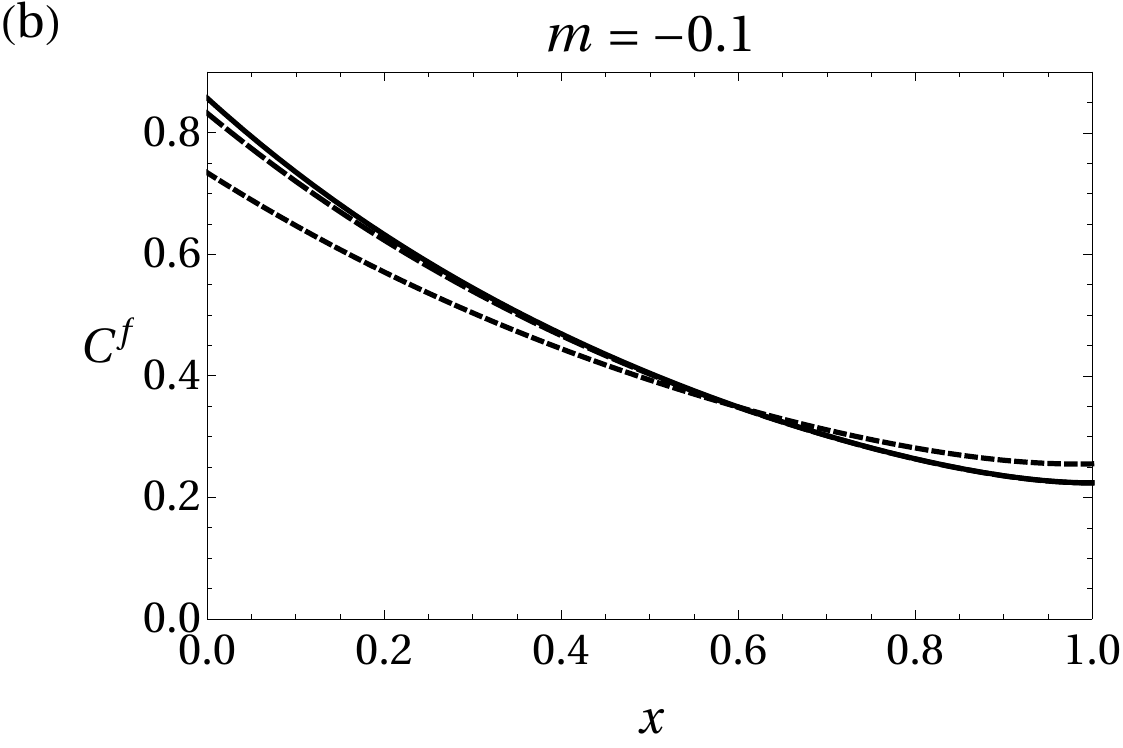}
	\caption{(a,b) Asymptotic and numerical concentration profiles for linear porosity given by \eqref{eq3.32}. The evaluation is performed for a two-dimensional square lattice with the parameters $\phi_0=0.75$, ${A=0.6,\, \mathrm{Pe}=3,\, k=1,\, m=-0.3}$ in (a) and $m=-0.1$ in (b).}
	\label{fig6}
      \end{figure}

With specified effective coefficients, we evaluate the asymptotic solution in Fig. \ref{fig6}. We see that the first-order approximation gives very good results in comparison with the numerical solution. We note that the deviation is slightly larger for the steeper gradient $m=-0.3$ in Fig. \ref{fig6}a, as should be expected because $\epsilon=m$. 

This close form solution allows further qualitative understanding of the filtration's dynamics. In particular, we may inspect the effect of key model parameters: $\mathrm{Pe}$, $A$ or the effect of porosity ($\phi_{0}$ or $m=\epsilon$), see Discussion and conclusions section.

\section{Filter efficiency} \label{sec5}
For a better understanding of the filtration efficiency, we analyse the obtained solutions a little further. We note that the optimisation question does not have a single goal as there might be several similar but distinct goals. For example, one might wish to minimise the contaminant concentration in the outflux from the filter or maximise the particle adsorption. However, as we have already discussed above, these are distinct goals (and dependent on $A$, that is the solute and solvent properties, filter gradient, fluid flow characteristics). Further one may wish to maximise the adsorption per unit time or per filter lifetime.

\begin{figure}
  \includegraphics[width=0.48\linewidth]{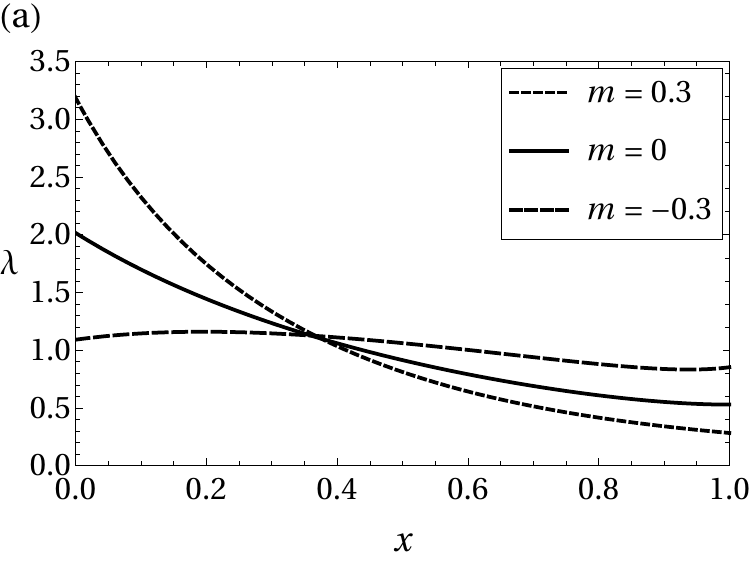}       	     \includegraphics[width=0.48\linewidth]{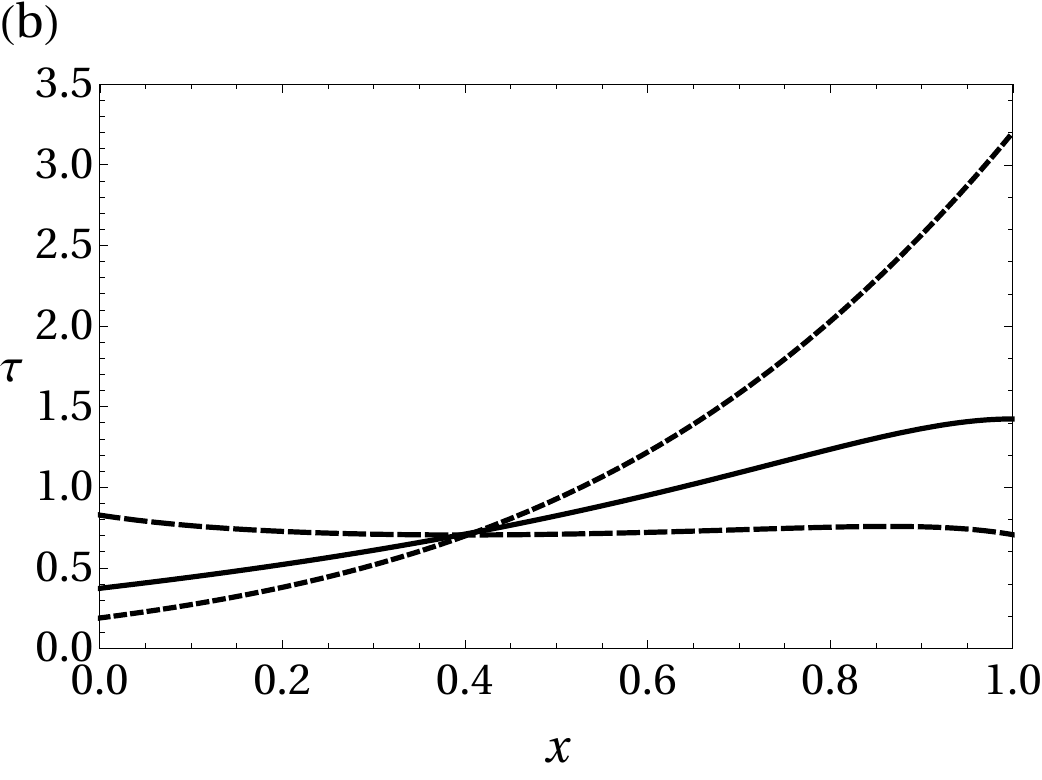}\\
	\includegraphics[width=0.48\linewidth]{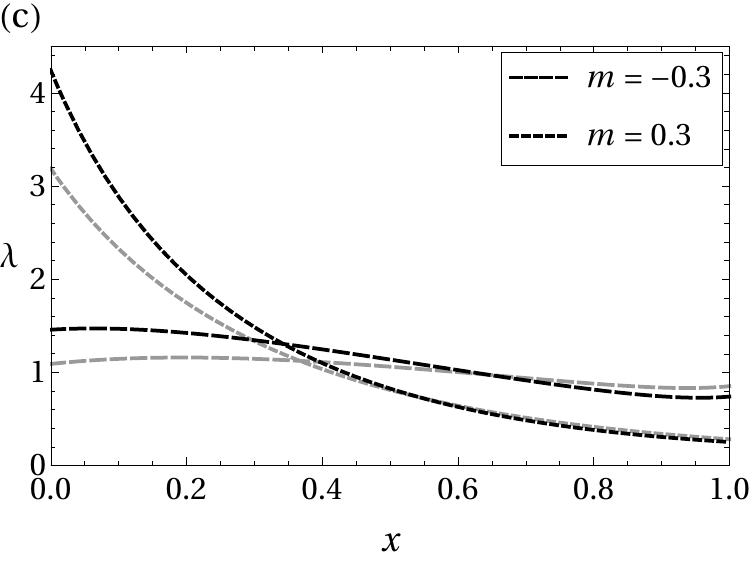} 
       \includegraphics[width=0.48\linewidth]{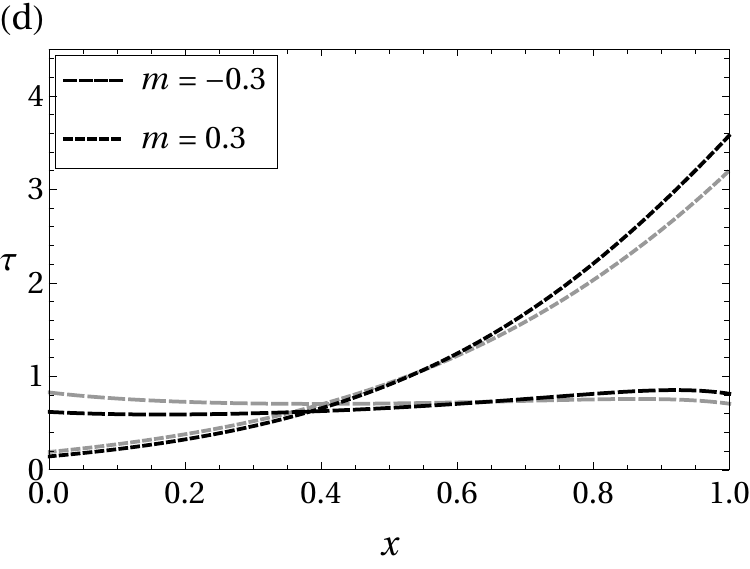}        
  \caption{ We plot two key profiles for efficiency characteristics for the case which was shown in Fig. \ref{fig3}. In panel (a) we show the adsorption profile $\lambda$ and in panel (b) the uniformity of time to local blockage of the filter for linear porosity \eqref{eq3.32}. A total measure of the former is in terms of $\Lambda$, while for the latter we introduced $\overline{T}$ in the text. In panels (c) and (d) we show the same characteristics but for hexagonal lattice, as in Fig. \ref{fig5}, where square lattice is in black curves while hexagonal lattice in grey.}
  \label{fig:3efficiency}
\end{figure}

We introduce several quantities measuring efficiency in various ways. First, we define a \emph{local adsorption rate} in the filter $\lambda(x)$ as
\begin{equation}  \label{eq.lambdameasure}
	\lambda(x)=f(\phi(x))C^f(x).
      \end{equation}
Note that this measure $\lambda$ has to be given in terms of the fluid-volume averaged concentration $C^{f}$ and not $C=\phi C^{f}$, see Eqn. \eqref{eq2.64f}. Indeed, $f C^f$ is concentration sink of the solute in a given fluid volume per unit time. On the contrary, $C$ depends the local amount of filter, via $\phi(x)$, and hence does not serve as a suitable measure of local adsorption rate.
      Since $\lambda$ is the right-hand side of \eqref{eq3.1a}, it represents the amount of filtered contaminant per unit area and unit time at the position $x$. The corresponding values of $\lambda$ for the concentration profiles depicted in Fig. \ref{fig3}a are shown in Fig. \ref{fig:3efficiency}a,c. We see that for the positive gradient $m=0.3$, the particle adsorption is significantly higher near the inlet of the filter than near the outlet. On the other hand, with a decreasing porosity gradient, the particle adsorption distribution becomes more uniform. This is consistent with the concentration change behaviour, which we discussed in Section \ref{sec3.2}.

Although it is useful to observe the particle adsorption distribution throughout the filter, it would be more appropriate to know the \emph{total adsorption rate by the whole filter per unit time}. This can be achieved by integrating the function $\lambda(x)$, which we define as a new metric $\Lambda$:
\begin{equation} \label{eq.Lambdameasure}
	\Lambda=\int_{0}^{1}\lambda(x)\dx.
\end{equation}

In our simple model of the filtration process, we simplify the filtration by completely removing the filtered out particles from the filter geometry. This assumption prevents the filter geometry from changing during the filtration process, which greatly simplifies the description. On the other hand, it also eliminates the possibility of filter clogging, which is in practice a very important property. Therefore, to fully analyse the filter efficiency, we need to estimate the clogging in a different way. To achieve that, we introduce a new metric
\begin{equation} \label{eq.Tmeasure}
	\overline{T}=\int_{0}^{1}\left|\tau(x)-\int_{0}^{1}\tau(s)\ds\right|\dx, \quad \tau(x)=\phi(x)/\lambda(x).
\end{equation}
describing \emph{the uniformity (in space) of time to local blockage of the filter}, noting that $\tau$           is proportional to time to local blockage of the filter, see Fig. \ref{fig:3efficiency}b,d. The smaller the value of $\overline{T}$ is, the more evenly is the contaminant adsorbed within the filter. If particle adsorption is non-uniform, the filter accumulates more of the contaminant in individual parts and the risk of clogging increases. In an ideal situation, we want the filter to clog all at once, rather than in a single location. And therefore, for the best efficiency when measured by the longevity of the filter, we seek the minimum value of $\overline{T}$.

As a final measure of filter efficiency, we suggest
\begin{equation} \label{eq4.41}
	R=\Lambda \min_{x\in[0,1]} \tau(x),
\end{equation}
being proportional to \emph{filtration capacity per filter lifetime} (the total amount of removed solute before clogging). This reflects the uniformity of adsorption (and hence filling up the pores) but with respect to fluid volume available. We do not aim to employ the detailed information about the microsctructure, hence the proposed measure $R$ is based on an estimate of fluid volume available via porosity.

We may now search for optimal designs of filter with linear gradients using these measures. However, note that further one needs to specify, which filtration scenario one wishes to optimise: with a given  \emph{i) fixed flow through the filter} or, \emph{ii) fixed  pressure drop across the filter}. The former simply utilises the decoupled fields $U$, $C^{f}$ to evaluate the measures, while the latter requires to assess pressure drop $\Delta p$ across the filter and rescale the model parameters so that the pressure drop remains fixed. Note that in order to calculate the decoupled pressure field, one needs solving the cell problem for $K$ which requires careful numerical approach, see Fig. \ref{fig2} and Appendix \ref{sec.App-K}.

\subsection{Fixed flow through the filter, case i)}

\begin{figure}
	\includegraphics[width=0.49\linewidth]{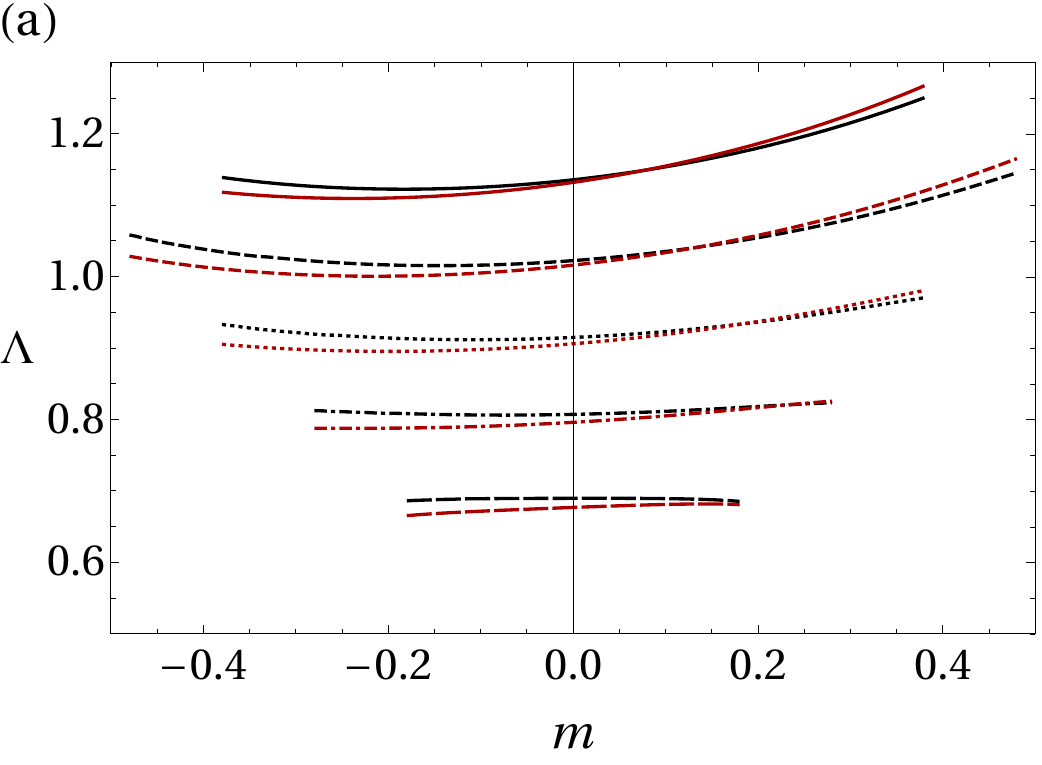}
	\includegraphics[width=0.49\linewidth]{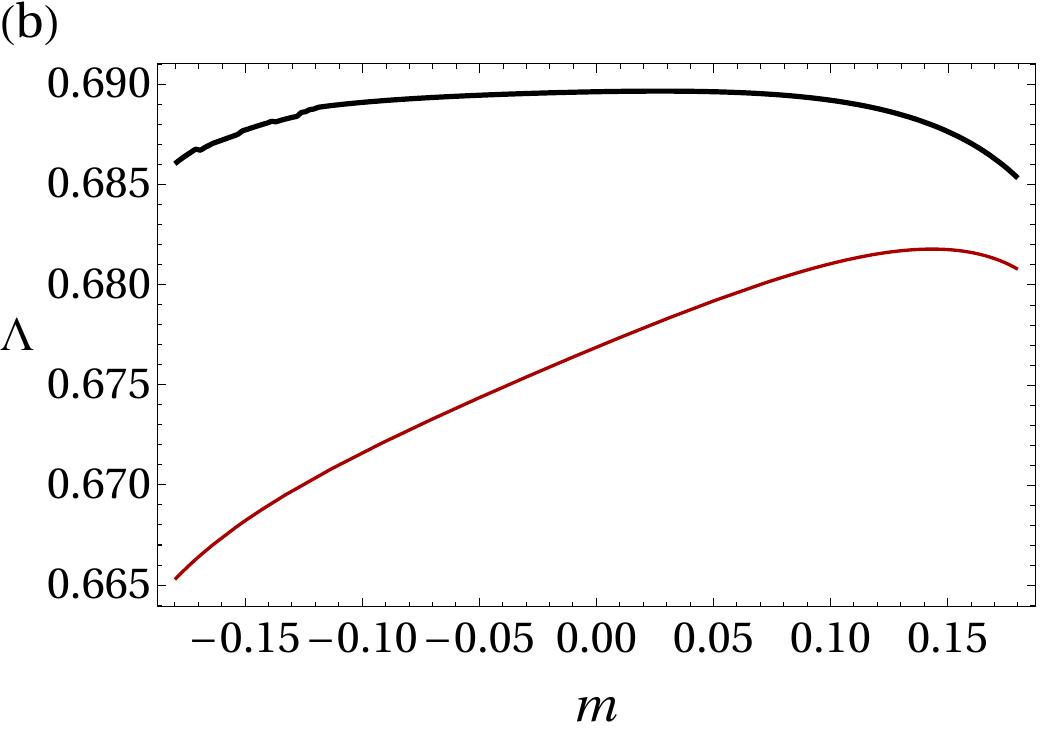} 
      	\includegraphics[width=0.49\linewidth]{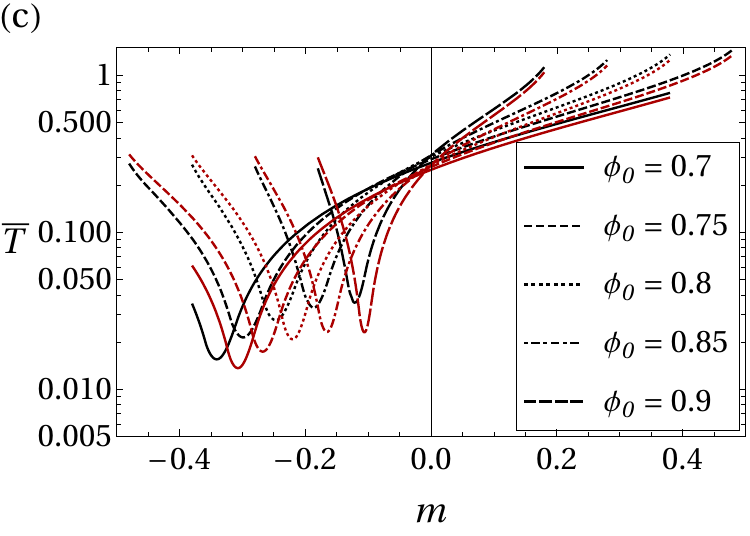}
        \includegraphics[width=0.49\linewidth]{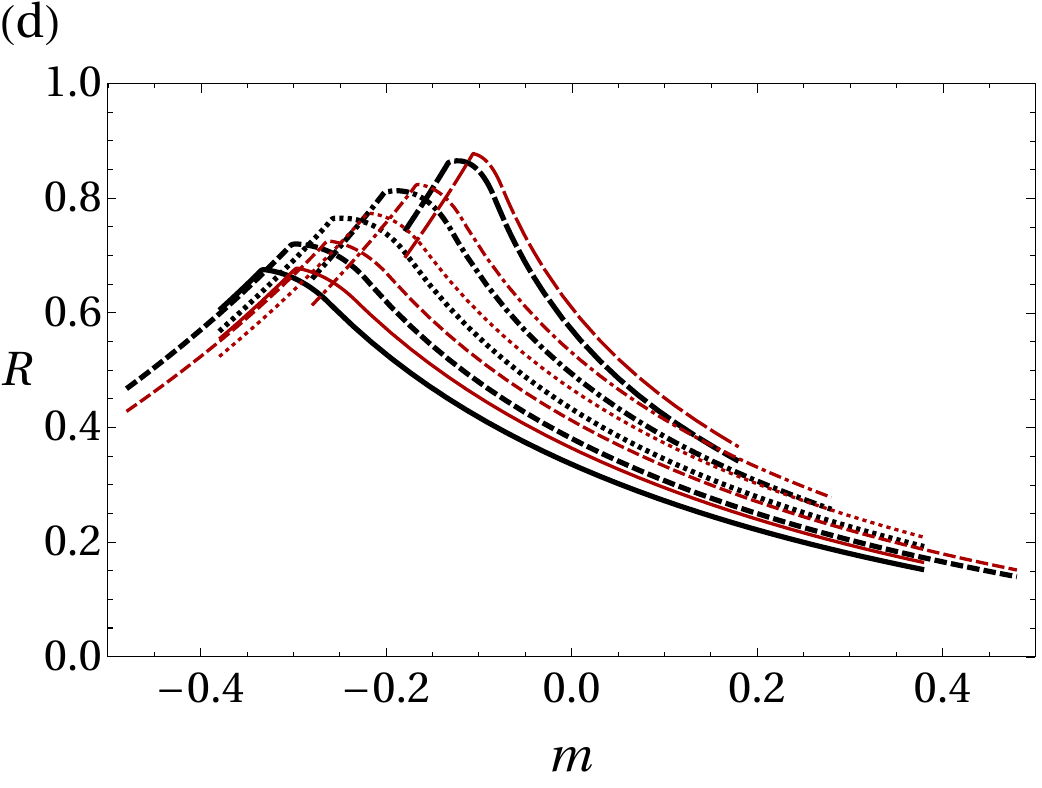}
	\caption{Numerically obtained (a two-dimensional square lattice with $A=0.6$, $\mathrm{Pe}=3$ and $k=1$) filter efficiency metrics: total adsorption rate per unit time $\Lambda$ (a,b), uniformity of time to local blockage $\overline{T}$ (c)), and filtration capacity per filter lifetime $R$ (d) for linear porosity \eqref{eq3.32} as a function of the porosity gradient $m$. Each curve represents a different mean porosity value $\phi_0$, starting from $\phi_0=0.7$ for the solid curve, up to $\phi_0=0.9$ (long dashed curve) with steps of $0.05$. At the same time, $m$ is continuously varied to observe the effect of filter design. Red curves correspond to $A=0$ (divergence free case), while black curves represent $A=0.6$ case. The variation in average porosity $\phi_{0}$ is depicted using different types of curve styles as shown in the boxed legend in panel (c). 
}
	\label{fig7}
      \end{figure}

We start by showing how the values of total adsorption rate $\Lambda$ depends on the filter design with linear gradient profile \eqref{eq3.32} for two-dimensional square lattice in Fig. \ref{fig7}a. For a fixed gradient, that is value of $m$, the value of $\Lambda$ increases with decreasing $\phi_0$. This is not surprising since a decreasing mean porosity $\phi_0$ results in a denser filter, allowing more of the solute to be filtered out. On the other hand, when fixing $\phi_0$ and varying the gradient $m$, we obtain more interesting results. First, the values of $\Lambda$ are asymmetric around the point $m=0$, showing that a positive gradient $m$ leads to a better filter efficiency in the sense of total particle removal per unit time. However, as shown in Fig. \ref{fig7}a,
the asymmetry seems to decrease with nonzero $A$ ($A=0.6$; black curves), which, in addition, seems to be generally improving $\Lambda$ characteristic for $m<0$, although the values for fixed $m$ are within $5\%$ difference. Note that the dependence of total adsorption rate $\Lambda$ highly depends on $\phi_{0}$ with lower values yielding more pronounced asymmetry and maximum being for largest positive slope $m$. Note however, that there is an exception for large porosity (in particular for $\phi_{0}=0.9$) where the best performance is for $m=0.027$, see Fig. \ref{fig7}b for more details and comparison with $A=0$.

When we inspect the filter efficiency from the perspective of performance in time, we obtain quite different results. Namely, high values of total adsorption rate $\Lambda$ are typically linked with a high degree of non-uniformity of removal within the filter as is illustrated by $\overline{T}$ plot in Fig. \ref{fig7}c. In view of those results, the uniformity of removal is significantly (by an order of magnitude) higher for negative porosity gradient $m$.

Similarly, the measure of filtration capacity per filter lifetime $R$ reaches maximal values for negative filter porosity gradients with optimal values similar but not the same as for the measure $\overline{T}$, see Fig. \ref{fig7}d.

Note that both measures $\overline{T}$ and $R$ show monotonous dependence on the average porosity $\phi_{0}$. The higher the average porosity is, the higher the filtration capacity per lifetime $R$ is but only for porosity gradients $m$ larger than the optimal value for a given averaged filter porosity $\phi_{0}$. For smaller $m$ there is an interval for $m$ such that this monotonicity is lost. In Fig. \ref{fig7}d this corresponds to the abrupt change in the plots of $R$ vs. $m$ which occurs due to the change in location of the filtration bottleneck (where the clogging event is expected to occur first). The uniformity of time to local blockage $\overline{T}$ also shows monotonic dependence on the average porosity $\phi_{0}$, but where there are essentially two regimes separated by $m\approx 0$. For smaller $m$, the larger $\phi_{0}$ is the smaller $\overline{T}$ is up to the optimal value (minimum of $\overline{T}$ with respect to $m$; below this optimal $m$ the monotonic dependence is lost). For $m>0$ we have the contrary, that is, the larger $\phi_{0}$ is, the larger $\overline{T}$ is.

In addition, we show the effect of nonzero velocity divergence (scaled by the value $A$) on these filter performance measures. Location of the optimal filter gradients is shifted in every studied case to lower values of $m$, higher negative porosity gradients, see Fig. \ref{fig7}c, d.




\subsection{Fixed pressure drop across the filter, case ii)}

Firstly, note that the decoupled system for $C,~\uvec$, Eqns.~\eqref{eq2.64}(a,b), is invariant under the scaling
\begin{equation*}
  (\uvec, \mathrm{Pe},k) \mapsto (\chi \uvec, \mathrm{Pe}/\chi, \chi k).
\end{equation*}
The decoupled equation for pressure, Eqn. \eqref{eq2.64c}, is in turn used to find a suitable scaling of the velocity field (including the inlet boundary condition $U(x=-\infty)$ for the unidirectionally graded filter) so that the pressure drop across the filter $\Delta p$ is fixed.


To reformulate the fixed (dimensional) pressure condition into a condition for $\chi$, we first choose a reference point, that is $U_{\mathrm{ref}}$ with a reference porosity gradient $\phi_{\mathrm{ref}}$ and fixed parameter values $\mathrm{Pe},~k$ (and nondimensionalisation constants $l, \mathcal{U}, \ldots$). The corresponding reference pressure drop satisfies
\begin{equation*}
  \Delta \tilde{p}_{\rf} = \frac{\mu}{(\delta l)^2} \int_0^{l} \frac{\mathcal{U} U_{\mathrm{ref}}(\x^*)}{K(\phi_{\mathrm{ref}}(\x^*))} \dx^*,
\end{equation*}
and hence we may write the condition of fixed pressure as
\begin{equation*}
    \Delta \tilde{p}_{\rf} = \frac{\mu}{(\delta l)^2} \int_0^{l} \frac{\mathcal{U} \chi U(\x^*)}{K(\phi(\x^*))} \dx^*.
  \end{equation*}
  Therefore, we have a simple relation for the sought scaling $\chi$ satisfying
  \begin{multline} \label{eq.FixedDeltaPcond}
    \chi = \int_0^{l} \frac{U_{\mathrm{ref}}(\x^*)}{K(\phi_{\mathrm{ref}}(\x^*))} \dx^* \left[\int_0^{l} \frac{ U(\x^*)}{K(\phi(\x^*))} \dx^*\right]^{-1}\\=\int_0^{1} \frac{U_{\mathrm{ref}}(\x)}{K(\phi_{\mathrm{ref}}(\x))} \dx \left[\int_0^{1} \frac{ U(\x)}{K(\phi(\x))} \dx\right]^{-1}  = \frac{\Delta p_{\rf}}{\Delta p}.
  \end{multline}
  Note, however, that as we use the fixed-flow solution with $U(-\infty)=1$ and rescaled $\mathrm{Pe},~k$, the parameter values have to be rescaled with inverse scaling $\chi^{{-1}}$ to reach the fixed reference pressure drop.

\begin{figure}
	\centering\includegraphics[width=0.65\linewidth]{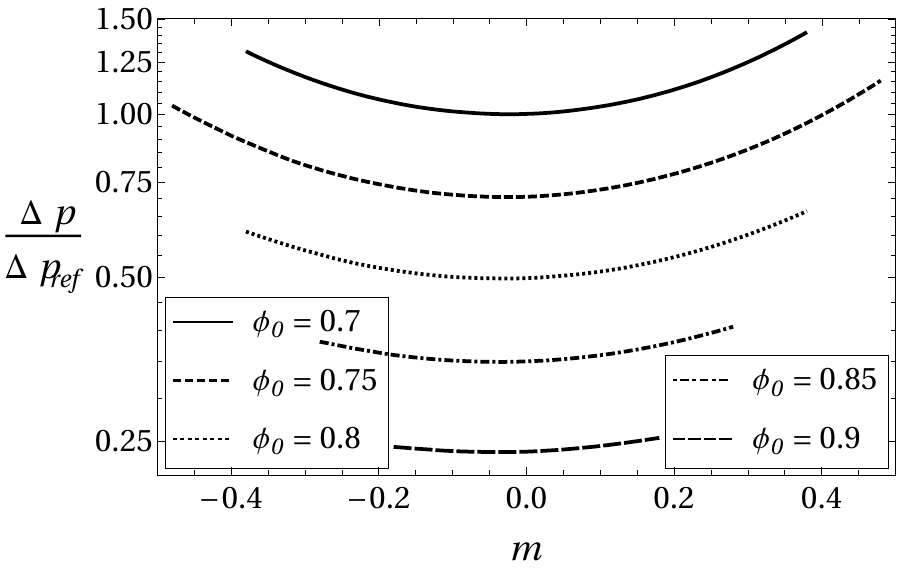}
	\caption{
         Numerical results for a two-dimensional square lattice with the parameters $A=0.6$, $\mathrm{Pe}=3$ and $k=1$ are shown. Changes in pressure drop across the filter in the unidirectional filter with linear porosity as a function of the filter design (that is the slope $m$ and its average porosity $\phi_{0}$). Note that we set $\Delta p_{\mathrm{ref}}$ so that it corresponds to the pressure drop with $\phi(x)=0.7-0.023(x-0.5)$, that is the minimum on the $0.7$ branch. (right panel) The dependence of pressure drop is nonlinear for $A\neq 0$ and hence assessing the appropriate scaling requires to resort to shooting method solving iterativelly the filter model equations.}
	\label{fig8}
      \end{figure}

  In the examples below, we choose the reference point for pressure $\Delta p_{\mathrm{ref}}$ as the minimum pressure drop on the $\phi_{0}=0.7$ branch. To see the variations in pressure drops with varying filter design  (still considering the linear gradients parametrised by the slope $m$ and the average value $\phi_{0}$), we plot $\Delta p/\Delta p_{{\mathrm{ref}}}$ for the considered filter designs (note that the minimum on branch $0.7$ is 1) in Fig. \ref{fig8}. 
 Note that even without invoking extreme values of permeability $K$, we can see one order of magnitude difference in pressure drops for fixed fluid flow and hence potentially significantly affecting the efficiency metrics. 

\begin{figure}
	\includegraphics[width=0.49\linewidth]{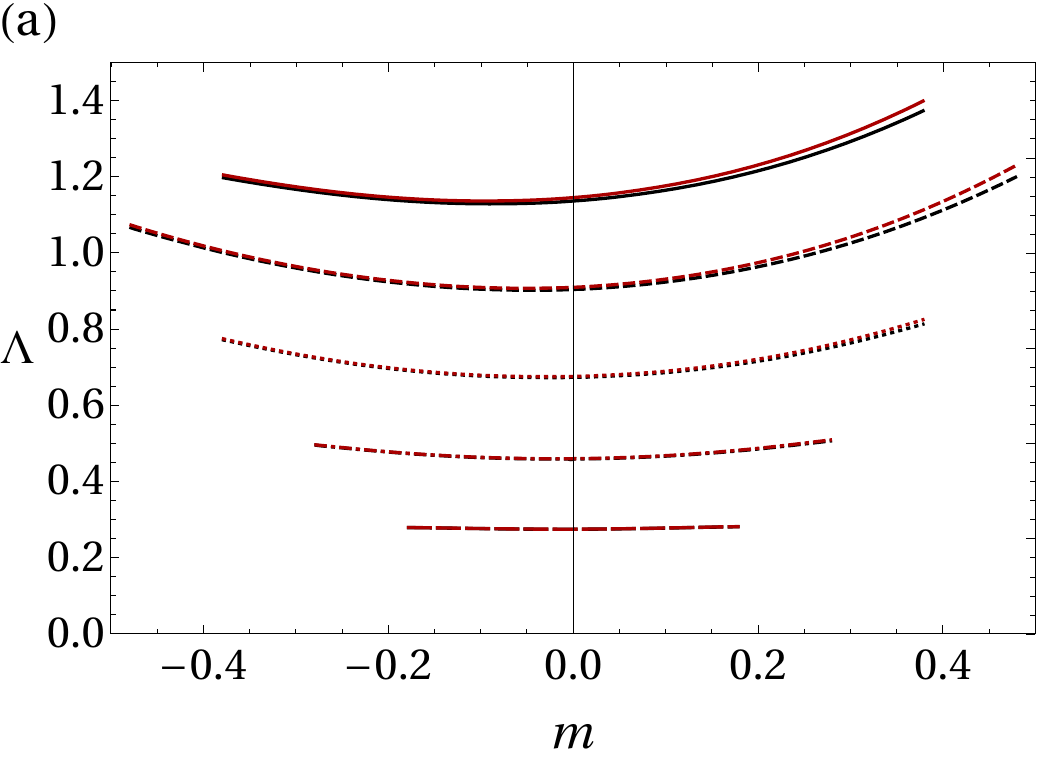}
	\includegraphics[width=0.49\linewidth]{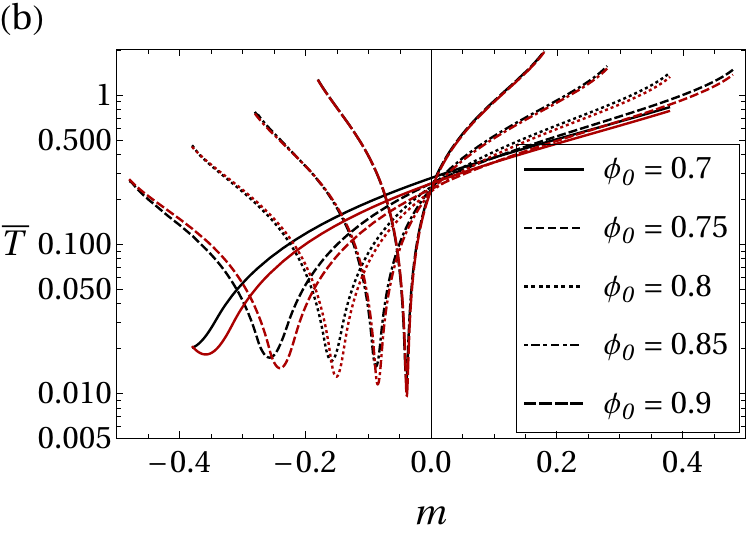}\\ 
\centering      	\includegraphics[width=0.49\linewidth]{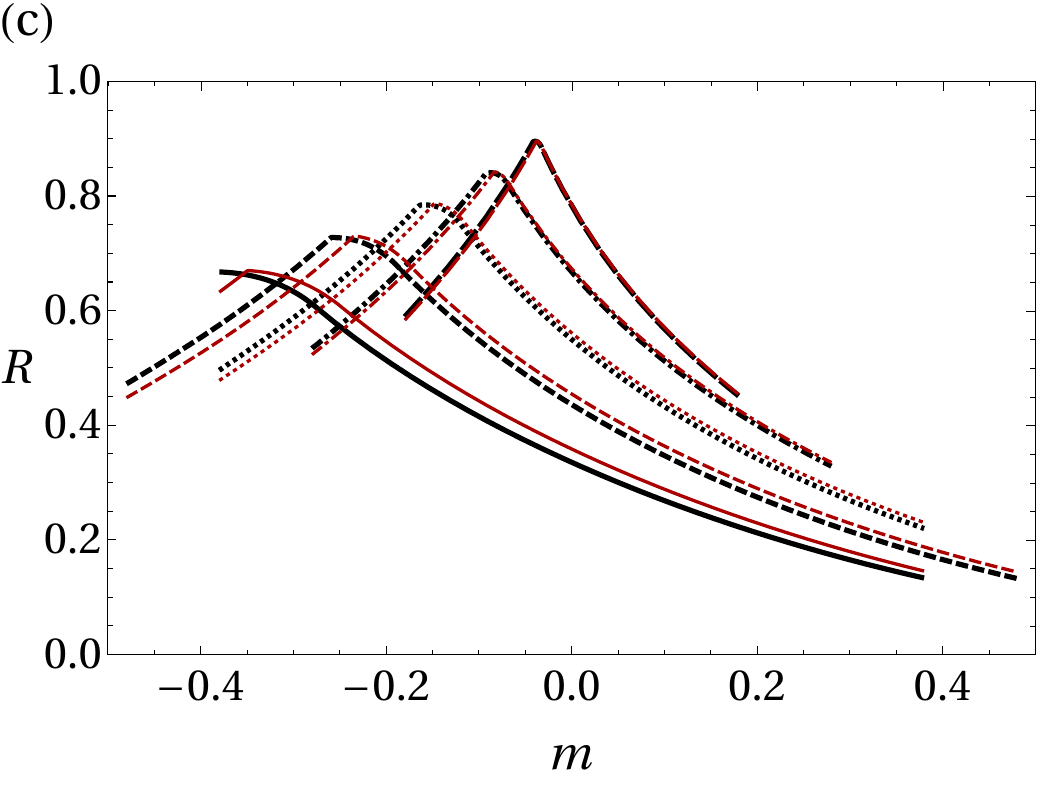}
	\caption{Numerically obtained (a two-dimensional square lattice with $A_{\mathrm{ref}}=0.6$, $\mathrm{Pe}_{\mathrm{ref}}=3$ and $k_{\mathrm{ref}}=1$) filter efficiency metrics  \emph{for fixed-pressure drop across the filter} (obtained by rescaling the fixed-flow problem): total adsorption rate per unit time $\Lambda$ (a), uniformity of time to local blockage $\overline{T}$ (b)), and filtration capacity per filter lifetime $R$ (c) for linear porosity \eqref{eq3.32} as a function of the porosity gradient $m$. Each curve represents a different mean porosity value $\phi_0$, starting from $\phi_0=0.7$ for the solid curve, up to $\phi_0=0.9$ (long dashed curve) with steps of $0.05$. At the same time, $m$ is continuously varied to observe the effect of filter design. Red curves correspond to $A=0$ (divergence free case), while black curves represent $A=0.6$ case. The variation in average porosity $\phi_{0}$ is depicted using different types of curve styles as shown in the boxed legend in panel (b). Note that the location of optimal values of $\overline{T}$ and $R$ are in a great agreement (with a maximum difference being less than $3\%$) suggesting their close relation. In particular, the maximas for $R$ and $A=0$ are: $-0.348,~-0.233,~-0.147,~-0.082,~-0.038$ while for $A=0.6$ we have: $-0.380,~-0.257,~-0.159,~-0.088,~-0.039$. Note that the significance of the mixture composition represented by $A\neq 0$ is increasing with decreasing porosity.   
}
	\label{fig9}
      \end{figure}

      We can see, that the difference in the capability of instantaneous filtration given in terms of $\Lambda$ is more pronounced in the fixed pressure drop when compared to the fixed-flow case, see Fig. \ref{fig7}a vs \ref{fig9}a. Further, the optimality of the total adsorption rate of the whole filter per unit time has shifted towards a more symmetric result with respect to negative and positive slope of the porosity gradient. However, the optimal filter design when viewed from the perspective of $\overline{T}$ and $R$ became more sensitive to the porosity profile. In particular, the optimal values of $m$ for various average porosities are more spread out. Finally, the significance of the nonzero $A$, that is the novel term resulting from the description of the filtration problem as a mixture entailing the nonzero divergence of velocity, is increased for smaller values of porosity, as expected.

 \section{Discussion and conclusions}

In this work, we have developed a generalised macroscopic model for filtration in graded porous media, accounting for non-solenoidal fluid velocity fields arising from solute-solvent coupling. The simplest performance metric considered is the outflux fluid-volume concentration, $C^{f}\!(1)$. While we proposed several measures, this quantity provides the most accessible insight into the influence of individual model parameters.

From the leading-order asymptotic solution, we observe that the concentration $C^{f}_{0}\!(1) = C_{0}(1)/\phi(1)$ depends exclusively on the leading-order diffusion coefficient $\tilde{D}_{0}$, the average porosity $\phi_{0}$, and the adsorption rate $f_{0}$. Consequently, in the limit of small porosity gradients, the leading-order filtration efficiency remains insensitive to both the specific filter design and the mixture composition (represented by the parameter $A$). In this regime, the process is governed primarily by the intrinsic transport and adsorption properties of the medium.

However, the first-order correction reflects the effects of both the composition of the mixture and the porosity gradient $m=\epsilon$. In particular, the correction $C_{1}$ depends on the modified flow field through $U_{1}$, which itself depends on $A$ (see Eqn. \eqref{eq4.66}). Through this coupling, the composition parameter influences the transport term and thus modifies the concentration profile and the resulting filtration efficiency. Consequently, even small porosity gradients introduce a weak but systematic dependence of the filtration efficiency on both the filter design and the mixture composition.

We also emphasise that different conclusions may arise depending on the chosen performance metric and on the experimental setup. Relaxing the assumption of small porosity gradients by the virtue of numerical solution, when considering the total adsorption rate $\Lambda$, we find that it is somewhat less than twice as large for the average porosity $\phi_{0}=0.7$ than for $\phi_{0}=0.9$. Moreover, in the former case the total adsorption rate is more sensitive to the porosity gradient, while the influence of the mixture composition ($A$) remains relatively small, typically on the order of a few percent. In the fixed--pressure drop configuration the situation changes: the total adsorption rate for $\phi_{0}=0.7$ remains comparable to that in the fixed--flow case, however it is approaching almost six times the value for $\phi_{0}=0.9$. In this regime the influence of the mixture composition appears nearly negligible within the considered range of porosities.

In contrast, the filtration capacity per lifetime, denoted by $R$, exhibits a different trend. In this case the performance is generally higher for larger porosity. Specifically, the peak value of $R$ for $\phi_{0}=0.9$ is approximately $20\%$ larger than that for $\phi_{0}=0.7$, while for a given porosity gradient $m$ it can be almost twice as large. For the fixed--pressure drop configuration the influence of the mixture composition becomes less pronounced, whereas the significance of the filter design on the overall performance increases. These observations illustrate that different optimisation objectives may favour different filter configurations and that performance measures such as $\Lambda$ and $R$ may lead to distinct design recommendations.

It is also instructive to consider the role of a nonzero coupling mixture parameter $A$. Recall that the flow velocity in the filter depends strongly on this parameter. In the special case $A=0$ (velocity field being solenoid) we have $U=1=U(-\infty)$, corresponding to a uniform velocity profile. When $A\neq 0$, however, the velocity field adjusts in response to the coupled mixture dynamics. In particular, we observe that the velocity at the entrance to the filter is typically reduced below the upstream value $U(-\infty)$, with the magnitude of this reduction depending on both the filter parameters and the mixture composition. This observation highlights the importance of imposing physically appropriate boundary conditions in the model that correctly reflect the experimental configuration. In practice, although the upstream flow may be prescribed, the velocity entering the filter is determined by the resistance of the porous medium and the coupled transport processes, and therefore need not coincide with the imposed upstream value.

Another way to interpret the parameter $A$ is that the filter performance depends on the composition of the mixture being filtered. In the fixed--flow configuration the resulting differences may reach $15$--$20\%$ for a given filter design (specified by $\phi_{0}$ and $m$), with both increases and decreases possible depending on the parameter regime. In the configuration corresponding to a fixed--pressure drop the effects of $A\neq 0$ appear less significant, although this is only partially true. Note that the Peclet number $\mathrm{Pe}$ increases with $\phi_{0}$ in this setup due to the rescaling associated with the imposed pressure drop, and therefore decreases as the pores gradually close. Consequently, the additional transport contribution arising from the full mixture description, which is proportional to $A/\mathrm{Pe}$, becomes increasingly important as the porosity decreases. This implies that even for arbitrarily small values of $A$ or $\mathrm{Pe}$, the contribution of this term eventually becomes non-negligible when the pores become sufficiently small.

Despite the improvements to the filtration model presented here, several limitations remain. While we have addressed the influence of graded porosity on transport, we have not explicitly accounted for the temporal evolution of the internal geometry due to particle accumulation. This phenomenon, known as filter blockage, has been explored in related contexts for periodic media through homogenised models of pore-scale narrowing \citep{dalwadi2016}. Incorporating such dynamic feedback into our non-solenoidal framework remains a subject for future research. Furthermore, the analytical insight into the resulting macroscopic filtration problem is inherently constrained by the complexity of the coupled governing equations. While the application of asymptotic methods allowed us to derive approximate solutions that reveal the physical significance of individual parameters, these results are not obtained in closed form and require numerical evaluation of the underlying cell problems.

Overall, our findings underscore the necessity of incorporating physically consistent mixture dynamics and boundary conditions in the modelling of porous filters. The interplay between filter architecture, operating conditions, and the coupling effects within the mixture provides a robust foundation for the future optimisation of high-efficiency graded filtration systems.

\begin{bmhead}[Declarations of competing interests.]
  Authors declare no conflict of interests.
\end{bmhead}


\begin{appen}

\section{Dimensionless form of governing equations} \label{App.sec2.2}
For easier manipulation with the derived systems \eqref{eq2.1} and \eqref{eq2.2}, we first convert them into a dimensionless form. To achieve this, we introduce new dimensionless variables $\x,\uvec,t,c$ and $p$ by relations
\begin{equation} \label{eq2.3}
	\tilde{\x}=l\x, \hspace{10pt} \tilde{\uvec}=\mathcal{U}\uvec, \hspace{10pt} \tilde{t}=\frac{l}{\mathcal{U}}t, \hspace{10pt} \tilde{c}=c_\infty c, \hspace{10pt} \tilde{p}=\left(\mu\frac{\mathcal{U}}{\delta^2 l}\right)p,
\end{equation}
 where $l$ is a characteristic length (typically the macroscopic size of the filter), $\mathcal{U}$ is a characteristic flow velocity and $c_\infty$ is a characteristic solute concentration, all of appropriate unit dimensions. We assume that the new variables $\x,\uvec,t,c$ and $p$ are all of order $\order(1)$ with respect to the small parameter $\delta$, introduced in Section \ref{sec2}. This parameter serves as a basis for asymptotic expansions made later, therefore all order comparisons are meant with respect to $\delta\to 0$. The pressure scaling factor $\frac{1}{\delta^2}$ is based on a observation that pressure in a capillary is proportional to the inverse square of its radius. Since the pore size in our model is proportional to the parameter $\delta\ll1$, we assume high pressure, specifically $\tilde{p}\sim\frac{1}{\delta^2}$ as follows from the typical scaling of permeability with cross-sectional area of a typical pore. Together with the transformation of the variables $\tilde{\x}$ and $\tilde{t}$, we also need to introduce new derivatives of the form
\begin{equation} \label{eq2.4}
	\begin{aligned}
		\pdv{}{\tilde{t}}&=\pdv{t}{\tilde{t}}\pdv{}{t}=\frac{\mathcal{U}}{l}\pdv{}{t}, \\
		\pdv{}{\tilde{x}_i}&=\pdv{x_i}{\tilde{x}_i}\pdv{}{x_i}=\frac{1}{l}\pdv{}{x_i} \implies \tilde{\bnabla}=\frac{1}{l}\bnabla.
	\end{aligned}
      \end{equation}


We now substitute the transformations \eqref{eq2.3} and \eqref{eq2.4} into the first set of equations \eqref{eq2.1}:
\begin{align*}
	\frac{\mathcal{U}}{l}\pdv{(c_\infty c)}{t}=\frac{1}{l}\bnabla\boldsymbol {\cdot}\left(\frac{D}{l}\bnabla(c_\infty c)-\mathcal{U}\boldsymbol{u}c_\infty c\right)&, \hspace{10pt} \x\in\Omega_f, \\
	-\gamma c_\infty c=\boldsymbol{n}\boldsymbol {\cdot}\left(\frac{D}{l}\bnabla(c_\infty c)-\mathcal{U}\boldsymbol{u}c_\infty c\right)&, \hspace{10pt} \x\in\partial\Omega_f,
\end{align*}
and after some algebra we obtain
\begin{subequations} \label{eq2.6}
	\begin{align}
		\pdv{c}{t}&=\bnabla\boldsymbol {\cdot}\left(\frac{D}{\mathcal{U}l}\bnabla c-\uvec c\right), \hspace{10pt} \x\in\Omega_f, \\
		-\frac{\gamma}{\mathcal{U}}c&=\boldsymbol{n}\boldsymbol {\cdot}\left(\frac{D}{\mathcal{U}l}\bnabla c-\uvec c\right), \hspace{11pt} \x\in\partial\Omega_f.
	\end{align}
\end{subequations}
Next, we define two new parameters: the Péclet number $\mathrm{Pe}=\frac{\mathcal{U}l}{D}$ and parameter $k=\frac{\gamma}{\delta \mathcal{U}}$, both assumed to be constant and of order \order(1). That is because we specifically focus on a situation where the adsorption coefficient $\gamma$ is small with respect to other quantities in \eqref{eq2.6}, i.e. $\gamma\sim\delta$. By this choice of parameters, we obtain a diffusion equation where the size of the individual terms is given only by the explicit appearance of the parameter $\delta$. After substitution of the parameters $\mathrm{Pe}$ and $k$ into \eqref{eq2.6}, we get a dimensionless form of the diffusion equation with boundary condition
\begin{subequations}
	\begin{align*}
		\pdv{c}{t}=\bnabla\boldsymbol {\cdot}\left(\frac{1}{\mathrm{Pe}}\bnabla c-\uvec c\right)&, \hspace{10pt} \x\in\Omega_f, \\
		-k\delta c=\boldsymbol{n}\boldsymbol {\cdot}\left(\frac{1}{\mathrm{Pe}}\bnabla c-\uvec c\right)&, \hspace{10pt} \x\in\partial\Omega_f.
	\end{align*}
\end{subequations}

We apply the same approach to the second set of equations \eqref{eq2.2}. First, we use the transformations \eqref{eq2.3} and \eqref{eq2.4}:
\begin{align*}
-\frac{1}{l}\bnabla\left(\mu\frac{\mathcal{U}}{\delta^2l}p\right)+\frac{\mu}{l^2}\bnabla^2\left(\mathcal{U}\uvec\right)=0&, \hspace{10pt} \x\in\Omega_f, \\
\frac{1}{l}\bnabla\boldsymbol {\cdot}(\mathcal{U}\uvec)=\left(\frac{1}{\rho_2^T}-\frac{1}{\rho_1^T}\right)\frac{1}{l}\bnabla\boldsymbol {\cdot}\left(\frac{D}{l}\bnabla(c_\infty c)\right)&, \hspace{10pt} \x\in\Omega_f, \\
	\mathcal{U}\uvec=0&, \hspace{10pt} \x\in\partial\Omega_f,
\end{align*}
and after few amendments, we obtain
\begin{subequations} \label{eq2.9}
	\begin{align}
		-\bnabla p+\delta^2\bnabla^2\uvec=0&, \hspace{10pt} \x\in\Omega_f, \\
		\bnabla\boldsymbol {\cdot}\uvec=\left(\frac{1}{\rho_2^T}-\frac{1}{\rho_1^T}\right)c_\infty \bnabla\boldsymbol {\cdot}\left(\frac{D}{\mathcal{U}l}\bnabla c\right)&, \hspace{10pt} \x\in\Omega_f, \\
		\uvec=0&, \hspace{10pt} \x\in\partial\Omega_f.
	\end{align}
\end{subequations}
We also define a new constant $A=\left(\frac{1}{\rho_2^T}-\frac{1}{\rho_1^T}\right)c_\infty$, which is apparently also dimensionless and of order \order(1). Let us also note that $A$ is generally nonzero, because we assume $\rho_2^T\neq\rho_1^T$ for our two-component mixture, as described in Section \ref{sec1.3.2}. Otherwise, we keep the constant $A$ general for now. After the substitution of the constants $A$ and $\mathrm{Pe}$ into \eqref{eq2.9}, we obtain a dimensionless set of equations for the fluid flow
\begin{subequations} 
	\begin{align*}
		-\bnabla p+\delta^2\bnabla^2\uvec=0&, \hspace{10pt} \x\in\Omega_f, \\
		\bnabla\boldsymbol {\cdot}\uvec=\frac{A}{\mathrm{Pe}}\bnabla^2 c&, \hspace{10pt} \x\in\Omega_f, \label{eq2.10b} \\
		\uvec=0&, \hspace{10pt} \x\in\partial\Omega_f.
	\end{align*}
\end{subequations}
Here the term $\frac{A}{\mathrm{Pe}}$ measures the deviation from the classical incompressibility condition as
described in Section \ref{sec1.3.2}.

\section{Details of permeability matrix $\K$ calculation} \label{sec.App-K}

First, note that the cell problem (the governing equations, boundary conditions and geometry) for $\K, \Piv$ is symmetric w.r.t. swapping  $1, 2$ in all indices. That is, the system posses symmetry: $$(y_{1},y_{2},K_{11},K_{12},K_{21},K_{22}) \leftrightarrow (y_{2},y_{1},K_{22},K_{21},K_{12},K_{11}),$$ and, as a result,
\begin{equation*}
  K_{22}(x,y)=K_{11}(y,x),\quad p_2(x,y)=p_1(y,x),\quad K_{21}(x,y)=K_{12}(y,x).
\end{equation*}
Therefore, the system of 6 unknowns car be rewritten into
\begin{subequations} \label{eq.App_3pdes}
\begin{align}
  1-\partial_{y_1} p_1(y_1,y_2) + \naby^2 K_{11}(y_1,y_2)&=0,\\
  -\partial_{y_2} p_1(y_2,y_1) + \naby^2 K_{12}(y_1,y_2)&=0,\\
  \partial_{y_1} K_{11}(y_1,y_2) + \partial_{y_1} K_{12}(y_2,y_1)&=0,
\end{align}
\end{subequations}
and
\begin{equation*}
 \int_{\Omega_f} K_{22}(x,y) \dy = \int_{\Omega_f} K_{11}(x,y) \dy.
\end{equation*}
Function $K_{12}$ is nonzero but its integral over $\Omega_{f}$ vanishes due to the symmetry of the problem, and therefore indeed
\begin{equation*}
  \K=K\Id=K_{11}\Id,
\end{equation*}
as stated in the main text, Eqn. \eqref{eq.Ksym}.

When implementing the above system of equations numerically, a great care has to be taken to avoid numerical blow up in the pressure relation. The issue lies in the compatibility of the numerical scheme with the property of the continuum system \eqref{eq.App_3pdes}. In particular, if we integrate the first and second equation over $\omega_{f}$ and employing the periodicity of all the fields, we get
\begin{equation} \label{eq.AppCompatibility}
  |\omega_{f}| = \int_{\partial \omega_{s}} \naby K_{11}\cdot \ny \dy,\quad 0=\int_{\partial \omega_{s}} \naby K_{12}\cdot \ny \dy.
\end{equation}
The former compatibility condition cannot be satisfied if there were no spherical obstacle or if there would be zero flux condition at the solid boundary. In other words, the periodicity of fields implies that there has to arise a nonzero flux of $K_{11}$ such that it counterbalances the term $|\omega_{f}|$. In addition, $p$ is determined up to a constant. Therefore, we implemented the cell problem for permeability tensor $\K$ and vector $\Piv$ in terms of equations \eqref{eq.App_3pdes} with finite differences, were the periodicity of all fields is directly enforced together with Dirichlet boundary conditions for $K_{{ij}}$ on $\partial \omega_{s}$, fixing $p_{1}$ free constant (setting the mean to zero) and explicitly enforcing the compatibility conditions via Lagrange multipliers. Only then we were able to acquire results in Fig. \ref{fig2}b noting that the implementation resulted in at least $10^5$ degrees of freedom (for the smallest porosity). The resulting linear problem was implemented and solved in MATLAB.

\end{appen}\clearpage

\bibliographystyle{jfm}
\bibliography{literatura}



\end{document}